\documentclass[lettersize,journal]{IEEEtran}

\usepackage[caption=false,font=footnotesize]{subfig}
\usepackage{cite}
\usepackage{url}

\usepackage{algorithm}
\usepackage{algorithmic}
\usepackage{multirow}
\usepackage{amssymb}
\makeatletter
\newcommand{\removelatexerror}{\let\@latex@error\@gobble}
\makeatother
\makeatletter
\long\def\@makecaption#1#2{%
  \ifx\@captype\@IEEEtablestring
    \footnotesize\bgroup\par\centering\@IEEEtabletopskipstrut
    {\normalfont\sffamily\footnotesize #1}\\{\normalfont\sffamily\footnotesize #2}\par\addvspace{0.5\baselineskip}\egroup\@IEEEtablecaptionsepspace
  \else
    \@IEEEfigurecaptionsepspace
    \setbox\@tempboxa\hbox{\normalfont\sffamily\footnotesize {#1.}\nobreakspace #2}%
    \ifdim\wd\@tempboxa>\hsize
      \parbox[t]{\hsize}{\normalfont\sffamily\footnotesize \centering {#1.}\nobreakspace #2}%
    \else
      \hbox to\hsize{\hfil\box\@tempboxa\hfil}%
    \fi
  \fi}
\makeatother
\usepackage{adjustbox} 
\usepackage{float}   
\usepackage{xcolor}  
\usepackage{listings}
\usepackage{seqsplit}

\usepackage{enumitem}
\usepackage[most]{tcolorbox}
\newtcolorbox{rqbox}{colback=gray!4,colframe=black!18,boxrule=0.4pt,
  arc=1mm,left=6pt,right=6pt,top=4pt,bottom=4pt}
\newcommand{\rqanswer}[2]{\begin{rqbox}\textbf{Answer to RQ#1.}~#2\end{rqbox}}
\usepackage[utf8]{inputenc}
\usepackage{listings-solidity}
\usepackage{xspace}
\usepackage{booktabs,tabularx}
\newcommand{\codename}{\mbox{\textsc{ObfProbe}}\xspace}
\newcommand{\hash}[1]{{\ttfamily\seqsplit{#1}}}

\definecolor{codegreen}{rgb}{0,0.6,0}
\definecolor{codegray}{rgb}{0.5,0.5,0.5}
\definecolor{codepurple}{rgb}{0.58,0,0.82}
\definecolor{backcolour}{rgb}{0.95,0.95,0.92}

\lstdefinestyle{mystyle}{
    backgroundcolor=\color{backcolour},   
    commentstyle=\color{codegreen},
    keywordstyle=\color{magenta},
    numberstyle=\tiny\color{codegray},
    stringstyle=\color{codepurple},
    basicstyle=\ttfamily\footnotesize,
    breakatwhitespace=false,         
    breaklines=true,                 
    captionpos=b,                    
    keepspaces=true,                 
    numbers=left,                    
    numbersep=5pt,                  
    showspaces=false,                
    showstringspaces=false,
    showtabs=false,                  
    tabsize=2
}
\definecolor{verylightgray}{rgb}{.97,.97,.97}

\title{Understanding and Characterizing Obfuscated Funds Transfers in Ethereum Smart Contracts}
\author{Zhang Sheng, TAN Kia Quang, Shen Wang, Shengchen Duan, Kai Li, Yue Duan%
\IEEEcompsocitemizethanks{%
\IEEEcompsocthanksitem Zhang Sheng, TAN Kia Quang, Shen Wang, Shengchen Duan, and Yue Duan are with Singapore Management University, Singapore (e-mail: dcszhang@foxmail.com; kq.tan.2023@msc.smu.edu.sg; shenwang918@gmail.com; sc.duan.2024@phdcs.smu.edu.sg; yueduan@smu.edu.sg).%
\IEEEcompsocthanksitem Kai Li is with Stevens Institute of Technology, Hoboken, NJ, USA (e-mail: kli50@stevens.edu).%
\IEEEcompsocthanksitem Corresponding authors: Kai Li and Yue Duan.%
}}

\begin{document}

\IEEEtitleabstractindextext{%
\begin{abstract}
    Scam contracts on Ethereum have rapidly evolved alongside the rise of DeFi and NFT ecosystems, utilizing increasingly complex code obfuscation techniques to avoid early detection. This paper systematically examines how obfuscation exacerbates the financial risks associated with fraudulent contracts and undermines existing auditing tools. We propose a transfer-centric obfuscation taxonomy, distilling seven strategies and seven measurable signals, and design \codename, a framework that performs bytecode-level smart contract analysis to uncover obfuscation techniques and quantify the level of obfuscation complexity via Z-score ranking. In a large-scale study of 1.04 million Ethereum contracts, we isolate over 3,000 highly-obfuscated contracts and, in manual case studies, identify four patterns: MEV bots, Ponzi schemes, fake decentralization, and extreme centralization that are deeply coupled with various obfuscation maneuvers, including assembly usage, dead code, and deep function splitting. We further reveal that obfuscation substantially increases the scale of financial damage and the time required for evasion. Finally, we evaluate SourceP, a state-of-the-art Ponzi detection tool, on both obfuscated and non-obfuscated samples, observing recall dropping from 0.80 to 0.12 and F1 from 0.89 to 0.21 under deep obfuscation in real-world scenarios. These findings underscore the urgent need for enhanced “anti-obfuscation” analysis techniques and broader community collaboration to mitigate the proliferation of scam contracts in the expanding DeFi ecosystem.
\end{abstract}
}
\maketitle
\IEEEdisplaynontitleabstractindextext

\section{Introduction}
\IEEEPARstart{S}{mart} contracts, which are self-executing programs deployed on blockchains, have been widely adopted recently. It has enabled various significant emerging applications, such as decentralized finance~\cite{compound:misc,burgerswap:misc,uniswap:misc} and digital art trading~\cite{Ruan2022}. Along with it, security issues are also on the rise. Recent reports~\cite{Wu2023,he2023detection} show that malicious smart contracts, including scams and MEV bots, have become increasingly prevalent, resulting in significant financial losses. To address these significant threats, researchers have proposed various techniques to detect~\cite{tann2018towards, sayeed2020smart, sathvik2024detection} and analyze~\cite{gupta2022deep, shah2023deep} these emerging security issues. Many of the techniques rely on static program analysis~\cite{ivanov2023security} and rule-based matching~\cite{agarwal2022vulnerability}, which have been proven highly effective and efficient in detecting early malicious smart contracts, typically straightforward and easily identifiable, such as transfers to externally owned private account addresses. 

As this arms race continues, malicious smart contracts are gradually replaced by more covert, complex, and obfuscated contract logic~\cite{Xia2025, Zhang2023, Sendner2023}. Recent studies~\cite {Xia2025, Zhang2023} confirm that obfuscation has become the primary means by which attackers conceal malicious transfer or backdoor control logic, which includes the use of assembly code, splitting functions, and redundant instructions. When attackers employ such obfuscation techniques in their smart contracts, the static analysis and matching rules adopted by traditional detection tools are often disrupted, leading to high detection inaccuracies~\cite{Sendner2023, Zhou2024} and further exacerbating financial losses~\cite{Zhou2023, Wang2024, Victor2022, Gupta2022}. Moreover, obfuscation techniques are not only found in malicious contracts but also in benign contracts for various reasons (e.g., to protect proprietary business logic and deter copycat attacks), which hinders security analysis and prevents users from better understanding the behaviors of the contracts. 

Although obfuscation techniques are increasingly employed in smart contracts and have been linked to substantial financial losses, no comprehensive study has yet systematically assessed their impact in real-world scenarios, an understanding that is critical for developing effective defense mechanisms and mitigating associated security risks. In this paper, we aim to conduct the first systematic study on the obfuscation of funds transfer operations, which are the most essential and security-critical activities of a smart contract~\cite{pan2023automated}. In particular, we aim to thoroughly understand the status quo of obfuscated funds transfer operations in Ethereum smart contracts by answering the following four essential research questions:
\begin{itemize}[leftmargin=*]
\item \textbf{RQ1 Definition:} What code obfuscation techniques are used for funds transfers in smart contracts, and how can they be defined and quantified?
\item \textbf{RQ2 Prevalence:} How prevalent are obfuscated funds transfers in the real world, and what is the current trend regarding the use of obfuscation techniques?
\item \textbf{RQ3 Financial Impact:} What are the consequences of using these techniques in malicious smart contracts with respect to economic impact? 
\item \textbf{RQ4 Impact on Malware Analysis:} Can state-of-the-art malicious contract detection tools maintain the same level of effectiveness when faced with heavily obfuscated funds transfers?
\end{itemize}

To answer these research questions, we develop a taxonomy to define and characterize obfuscation techniques in fund transfer operations. Specifically, we propose seven robust features by systematically dissecting and examining the formal definition of funds transfer operations within the Ethereum virtual machine to ensure the comprehensiveness and the representativeness of the feature list. We quantify the obfuscation complexity of fund transfer operations in smart contracts using a Z-score-based representation model. We then propose \codename, an EVM bytecode analysis framework that accurately uncovers various obfuscation techniques used in real-world smart contracts. Using this analysis framework, we conduct a series of studies on real-world smart contracts to examine the use of obfuscation techniques for both malicious and benign purposes in funds transfer operations. Below, we highlight some interesting discoveries:
\begin{itemize}[leftmargin=*]
\item[(1)] By analyzing the top 3000 highly obfuscated contracts detected by \codename, we found 463 contracts that exhibit substantial security risks, placing funds totaling $\approx$\$100 million at risk.

\item[(2)] Compared to non-obfuscated scam contracts, obfuscated scam contracts demonstrate a significantly larger financial impact, with their highest recorded inbound funds being $\approx$2.4X higher and clear periods of intensified victimization occurring between 2018 and 2025.

\item[(3)] The evaluation of a state-of-the-art Ponzi detector on obfuscated Ponzi contracts shows a significant effectiveness downgrade (e.g., F1 score from 0.89 to 0.21), demonstrating that obfuscations can significantly undermine the performance of existing detection tools.
\end{itemize}


\vspace{4pt}\noindent \textbf{Contributions.} The contributions of this paper are summarized as follows:
\begin{itemize}[leftmargin=*]
\item We developed \codename, the first EVM bytecode obfuscation analyzer that leverages seven bytecode-level features and a Z-score representation model to automatically detect obfuscated transfer logic from smart contracts.

\item We conducted a large-scale measurement of more than 1.04 million Ethereum smart contracts, revealing common obfuscation patterns in the top 3000 contracts, including four types: MEV bots, Ponzi schemes, fake decentralization, and extreme centralization.

\item We quantitatively studied the impact of obfuscation on financial damage and detection.
\item We open-sourced \codename and the collected data/artifacts to facilitate future research\footnote{https://github.com/nonname-byte/Obfuscation\_Tool}.
\end{itemize}


\section{Background}
\label{background}
\subsection{Blockchain and Smart Contracts}
Blockchain, a technology with a decentralized distributed ledger at its core, ensures data security and immutability through cryptographic methods, making it a foundational infrastructure for various data transactions~\cite{ali2021ensuring}. Its core features—decentralization, transparency, immutability, and security—position it as a transformative technology across a wide range of industries~\cite{tariq2019blockchain}.

Smart contracts are programs built on top of blockchains that autonomously execute contractual terms when predefined conditions are satisfied, eliminating the need for intermediaries~\cite{rashid2019smart}. Smart contracts offer several key advantages, including reduced transaction costs, enhanced efficiency, and a lower risk of human error. By permanently recording execution results and data on the blockchain, they eliminate the possibility of unilateral alteration, thereby ensuring fairness and transparency in contract execution~\cite{sakib2024blockchain}. Smart contracts have been widely applied in various fields, including financial transactions, identity verification, supply chain management, and insurance~\cite{sakib2024blockchain,tariq2019blockchain}. In the financial sector, smart contracts enable the automatic settlement and payment on the blockchain, significantly enhancing transaction efficiency and transparency while reducing the need for intermediaries and human intervention. As blockchain technology evolves, smart contracts will play an increasingly important role across multiple industries~\cite{tariq2019blockchain}.

\subsection{Code Obfuscations}
\label{sec:back:obfuscation}
Code obfuscation transforms a program into a semantics-preserving but harder-to-analyze form, used both for software protection and for malware evasion to raise reverse-engineering cost. A standard taxonomy groups techniques into control-flow, data, layout (lexical), and instruction-substitution transformations \cite{Collberg1997Taxonomy}. Control-flow obfuscation perturbs the CFG (e.g., opaque predicates, bogus branches, flattening) and even virtualization to hinder static/dynamic analyses \cite{Li2022,Schloegel2022}. Data obfuscation hides computations via variable encoding and mixed Boolean–arithmetic (MBA) rewriting; layout changes rename/reorder identifiers or inject dead/NOP code; instruction substitution replaces code with equivalent forms (e.g., shifts/adds for multiply), which diversifies byte patterns and weakens signature-based detection \cite{Little2023,Collberg1997Taxonomy,Schloegel2022}.

\section{Taxonomy of Obfuscated Funds Transfers}

Our study focuses on funds transfer operations, which are the most essential and security-critical activities of a smart contract~\cite{pan2023automated}. To answer our research questions, we develop a taxonomy to define and characterize different obfuscation techniques on funds transfer operations, based on how each component of funds transfer operations can be hidden, derived from the formal definition of funds transfer operations within the Ethereum virtual machine to ensure the comprehensiveness and the representativeness of our taxonomy.

We start by dissecting every element of the funds transfer operation. The standard way to realize such an operation is through a \textit{transfer} API CALL in Solidity~\cite{common:misc}, which is implemented as a \textit{CALL} EVM opcode in a given smart contract bytecode. Specifically, we define a \textit{CALL} EVM opcode that implements a funds transfer operation as follows:

{\bf Definition 1.} A \textit{CALL} EVM opcode that implements a funds transfer operation can be defined as

\vspace{-0.15in}
\begin{equation}\label{eq1} 
T = (addr, value, context, log), where: 
\vspace{-0.05in}
\end{equation}

$\bullet$ \textit{addr} determines the recipient address of the fund. This is the target address specified in the \texttt{CALL} instruction (a 20-byte Ethereum address).

$\bullet$ \textit{value} is the non-zero value in wei transferred to the \texttt{addr}. This value represents the amount of native Ethereum tokens sent in the transfer.

$\bullet$ \textit{context} is the execution context of the funds transfer operation, which includes the storage state of the contract, the remaining instructions, and control/data flow inside the function where the transfer is located. This information together determines whether and how the \texttt{CALL} instruction can be executed.

$\bullet$ \textit{log} refers to the collection of events generated by the \texttt{CALL} operation during the transaction execution. This includes event signatures (topics) and data fields (data), which provide semantic information regarding the transfer and are useful for auditing and monitoring on-chain activities.

Based on our definition of a funds transfer operation, we investigate each element and derive seven obfuscation features. By exhaustively mapping each feature to one or more elements of a funds transfer operation, we ensure our taxonomy is comprehensive and covers all fundamental methods of hiding transfer operations in Ethereum smart contracts. Please note that our analysis is conducted at the bytecode level, and the source code listings below are provided for illustrative purposes.


\subsection{Obfuscation of \texttt{addr}}
The \texttt{addr} element is essentially a string that represents a 20-byte Ethereum address, indicating the recipient address of a funds transfer operation. We consider four obfuscation methods that stem from traditional string obfuscation techniques~\cite{quiring2019misleading}.

\vspace{4pt}\noindent \textbf{T1. Multi-step Address Generation.} The address is derived through a sequence of external reads, arithmetic/bitwise operations, or import from another contract, preventing straightforward identification of the actual 20-byte recipient.

\begin{lstlisting}[language=Solidity,caption=T1 Multi-step Address Generation]
// Step 1: derive seed from block data
bytes32 seed = keccak256(...);
// Step 2: extract intermediate bytes
bytes20 part = bytes20(seed);
// Step n......
// compute recipient address
address rec = address(...(uint256(part)));
// core transfer
rec.transfer{value,...}("");
\end{lstlisting}

\vspace{4pt}\noindent \textbf{T2. Complex String Operations.} The address is split into multiple substrings or byte segments stored separately. These segments are then concatenated at runtime to reconstruct the true \texttt{addr}, concealing it from static parsers.
\begin{lstlisting}[language=Solidity,caption=T2 Complex String Operations]
// split address string into parts
string memory s1 = "0x";
string memory s2 = "a1b2c3";
string memory s3 = "d4e5f6";
// concatenate at runtime
string memory full = string(s1, s2, s3);
// parse back to address
address rec = parseAddr(full);
rec.transfer{value,...}("");
\end{lstlisting}

\vspace{4pt}\noindent \textbf{T3. External Contract Calls.} Instead of local computation, the contract with this obfuscation technique may choose to query a “router” or “delegate” contract to fetch \texttt{addr}, hiding the true recipient behind an external CALL.
\begin{lstlisting}[language=Solidity,caption=T3 External Contract Calls]
interface AddrPro {
    function getAddr() external returns (address);
}
// fetch hidden address from another contract
address provider = 0x1234...;
address rec = AddrPro(provider).getAddr();
rec.transfer{value: value}("");
\end{lstlisting}

\vspace{4pt}\noindent \textbf{T4. Control-flow complexity.}
The branch selection is dependent on run-time conditions (e.g., \texttt{block.timestamp}). Some branches are dummy and never execute, and different branches point to different \texttt{addr}s. This hides the real recipient because the true branch is known only at execution time, which hinders static rule matching.
\begin{lstlisting}[language=Solidity,caption=T4 Control-flow Complexity]
address rec;
if (block.timestamp % 3 == 0) {
    for (uint i = 0; i < 2; i++) {
        if (i == 1) {
            if (msg.sender == owner)
                rec = addrA;
            else
                rec = addrB;
        }
    }
} else
    rec = addrC;
rec.transfer{value: value}("");
\end{lstlisting}

\subsection{Obfuscation of \texttt{value}}
Since \texttt{value} is the amount transferred in wei, i.e., a 256-bit unsigned integer. Therefore, two obfuscation strategies from \texttt{addr} can also be applied.

\vspace{4pt}\noindent \textbf{T3. External Contract Calls.} Similar to \texttt{addr}, the transfer amount can also be fetched from an external contract, rather than stored locally, complicating static quantity analysis.

\vspace{4pt}\noindent \textbf{T4. Control‐flow complexity.} \texttt{value} is determined by conditional logic or loops, introducing multiple potential numbers and obscuring the true transfer amount.

\subsection{Obfuscation of \texttt{context}} 
The \texttt{context} element comprises (i) the contract’s storage state, (ii) the internal instructions and control/data flow of the function in which the funds transfer operation resides. Smart contract developers can choose to obfuscate \texttt{context} to cloak the real intent of funds transfer operations. We outline two unique techniques below to obfuscate \texttt{context}.

\vspace{4pt}\noindent \textbf{T5. Camouflage Instructions.} By injecting large numbers of meaningless loops, arithmetic operations, and \texttt{NOP}s into the transfer-related function body, the core \texttt{CALL} is camouflaged with many irrelevant instructions. Although the transfer is still executed correctly, the altered control and data flow within the function make it hard for static analyzers to isolate the actual context.

\begin{lstlisting}[language=Solidity,caption=T5 Camouflage Instructions]
// meaningless loop
for (uint i = 0; i < 5; i++)
    uint tmp = i * 42;
// no-op arithmetic
uint x = (1 + 2) - 3;
// core transfer hidden among noise
address rec = 0xAbCd...;
rec.transfer{value,...}("");
\end{lstlisting}

\vspace{4pt}\noindent \textbf{T6. Replicated Transfer Logic.} Smart contract developers can duplicate identical or highly similar transfer logics across multiple functions that only differ in names (e.g., \texttt{withdraw} and \texttt{start}) or trivial execution paths. Hence, the contract selector randomly dispatches the CALL at runtime. This multiplies potential entry points and confuses analysis tools regarding which function’s context truly carries out the transfer.
\begin{lstlisting}[language=Solidity,caption=T6 Replicated Transfer Logic]
function withdraw(uint value) public {
    _doTransfer(value);
}
function start(uint value) public {
    _doTransfer(value);
}
function _doTransfer(uint value) internal {
    // same transfer code reused
    address rec = 0xAbCd...;
    rec.transfer{value,...}("");
}
\end{lstlisting}

\subsection{Obfuscation of \texttt{log}}
Finally, \texttt{log} keeps a record and provides a semantic-level understanding of what has happened during the execution. Developers can choose to obfuscate the semantic signals in \texttt{Log} with the following technique.

\vspace{4pt}\noindent \textbf{T7. Irrelevant Log Events.} One can emit misleading or unrelated events (e.g., logging a transfer to a “legitimate” address while sending funds elsewhere), diverting auditors’ and tools’ attention, and concealing the real log data that corresponds to the transfer.
\begin{lstlisting}[language=Solidity,caption=T7 Irrelevant Log Events]
event Info(string msg);
// misleading log before transfer
emit Info("Sending to safe address");
address rec = 0xAbCd...;//unsafe address
rec.transfer{value,...}("");
\end{lstlisting}

\rqanswer{1}{We answer RQ1 by formalizing the EVM transfer path, deriving seven obfuscation \emph{patterns} (T1--T7) and mapping each to measurable bytecode features (F1--F7). This taxonomy operationalizes transfer obfuscation and provides the quantitative basis used by our Z-score model in Section~\ref{System}.}

\section{System Design and Implementation}
\label{System} 

Building on the taxonomy, we design and implement an EVM bytecode analysis tool named \codename to uncover how different obfuscation techniques manifest in real-world smart contracts, thereby enabling our RQ2 measurement.

\subsection{System Overview}

\begin{figure*}[h]
    \centering
    \includegraphics[width=0.8\linewidth]{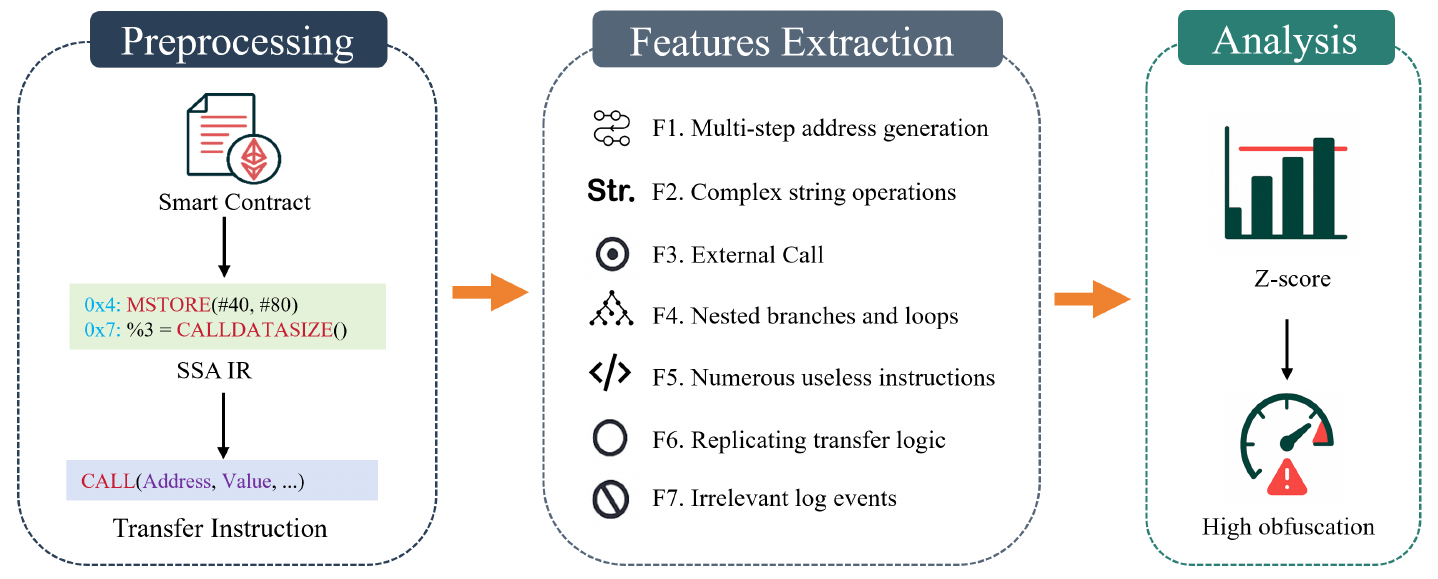}
    \caption{Overview of the analysis pipeline of \codename.}
    \label{fig:tool-pipeline}
\end{figure*}
Figure~\ref{fig:tool-pipeline} shows an overview of \codename. At a high level, it converts a given smart contract bytecode to the static single assignment intermediate representation (SSA IR) using an existing tool named Rattle~\cite{crytic2025rattle}. After that, it scans the IR to detect all fund transfer operations. For each transfer, our system extracts seven pre-defined obfuscation features. Finally, it applies a Z-score representation model~\cite{curtis2016mystery}, which is a numerical value that represents a data point's distance from the mean in terms of standard deviations, to convert the extracted features into an obfuscation score that indicates the degree of complexity in the obfuscation applied to the transfer operation.

\begin{table*}[h]
\centering
\small
\caption{Summary of transfer-related obfuscation features.}
\label{tab:feature_summary}
\begin{tabularx}{\textwidth}{@{} l X @{}}
\toprule
\textbf{Obfuscation strategies} & \textbf{Extracted features} \\
\midrule
T1. Multi-step address generation & F1. Number of steps in the \texttt{addr} generation. \\
T2. Complex string operations     & F2. Number of string operations in the \texttt{addr} generation. \\
T3. External contract calls       & F3. Presence of an external contract call in the \texttt{addr}/\texttt{value} derivation. \\
T4. Control-flow complexity       & F4. Max branch/loop nesting depth along \texttt{addr}/\texttt{value} derivation. \\
T5. Camouflage instructions       & F5. Transfer-related Instruction Ratio (TIR). \\
T6. Replicated transfer logic     & F6. Inter-function similarity among transfer-containing functions. \\
T7. Irrelevant log events         & F7. Semantic relevance between log events and the transfer operation. \\
\bottomrule
\end{tabularx}
\vspace{-0.15in}
\end{table*}

\subsection{Definition and Extraction of Obfuscation Features}
\label{Quantifying Obfuscation via Transfer Features}

For each of the aforementioned seven obfuscation strategies, we define a corresponding obfuscation feature that can be extracted by \codename, as summarized in Table~\ref{tab:feature_summary}.

\vspace{4pt}\noindent \textbf{F1. Number of steps in address generation.} This numerical feature represents the number of steps required to obtain \texttt{addr} in a transfer operation. Starting from each \texttt{CALL} operation in the contract's SSA IR, we perform backward dataflow analysis on the parameter \texttt{address} to trace its generation process. Each arithmetic operation, hash function invocation, bitwise manipulation, and external call is counted as one distinct step. Finally, we consolidate linear operations within each basic block to avoid overcounting trivial operations, and we treat the number of steps for address generation as a numerical feature F1. 

\vspace{4pt}\noindent \textbf{F2. Number of string operations.} String operations (e.g., concatenation, hashing, slicing, and encoding) contribute to obfuscation. To quantify the complexity of string operations in the address generation process, we reanalyze the data flow of the parameter \texttt{addr} to count all instructions that involve string manipulation, including both built-in string operations and hash operations. We then count the number of string operations for address generation as a numerical feature F2.

\vspace{4pt}\noindent \textbf{F3. Presence of an external call.} To determine if the generations of \texttt{addr} and \texttt{value} involve any external call, we inspect the dataflow of both and check if there exists any external call (i.e.,\texttt{CALL}), we set a binary feature F3 to \texttt{TRUE}, otherwise, to \texttt{FALSE}.

\vspace{4pt}\noindent \textbf{F4. Height of the branch tree.} We use the height of the branch tree as a numerical feature to represent the complexity of the control flow associated with the transfer operation by analyzing conditional branch structures (\texttt{JUMPI} instructions). Starting from the transfer operation, we backtrace to traverse all conditional branches. The height of the branch tree is calculated as the maximum nesting depth traversed.

\vspace{4pt}\noindent \textbf{F5. Transfer-related instruction ratio (TIR).} This feature quantifies the ratio of effective instructions contributing to a transfer operation in the residing function. To this end, we define \emph{Transfer-related Instruction Ratio} (TIR) as:  $\text{TIR} = \frac{|U|}{|T|}$, where $|U|$ is the number of \emph{effective} instructions contributed to transfers and critical state updates, and $|T|$ is the total number of instructions in the transfer-residing function. Specifically, \(U\) is computed as the union of critical transfer instructions \(C\) and state modifications \(S\).  

We obtain \(C\) from the function's program dependence graph (PDG) by considering all instructions directly related to the transfer logic (e.g., \texttt{CALL}, parameter calculations, state validation). \(S\) is from parsing the function to detect \texttt{SSTORE} operations and retaining only those updating parameters related to transfers (e.g., \texttt{address}, \texttt{value}). Consequently, a low TIR indicates that a large portion of the code in a transfer-resident function does not affect how the transfer operation is performed, suggesting significant obfuscation designed to dilute and hide the contract's true transfer logic. 


\vspace{4pt}\noindent \textbf{F6. Transfer operation similarity.} This feature quantifies whether a contract replicates the same transfer logic across multiple functions. To calculate this, we construct a PDG that includes information from both control and data dependencies for each function containing a transfer operation. Then, we embed the nodes of each PDG using a Word2Vec model and further employ a Relational Graph Convolutional Network (R-GCN)~\cite{schlichtkrull2018modeling} to aggregate the node embeddings and edge relations into a single vector representation for each function’s subgraph. Finally, we compute pairwise cosine similarities between these vector representations across all transfer-containing functions and use these scores as numerical feature F6. A larger similarity score indicates that two transfer-containing functions implement more similar transfer logic. We treat this feature as a continuous indicator rather than a binary decision, allowing for some noise in the functions. 

\vspace{4pt}\noindent \textbf{F7. Relevance of log events.}
F7 captures whether log events emitted along transfer logic are \emph{semantically unrelated} to the transfer itself (i.e., noisy or misleading logs). We decode log topics/data when ABI signatures are available and semantically classify each log as transfer-related or not using an LLM-assisted classifier. Our design includes a rollback mechanism that resolves ambiguous cases (i.e., those with low LLM confidence scores) via manual annotation. In a 100-sample pilot, the LLM achieves 100\% accuracy on this binary relevance task, so we use it to support semantic labeling. We set F7 to \texttt{TRUE} if irrelevant logs co-occur with transfer operations, otherwise \texttt{FALSE}.

\subsection{Obfuscation Z-score Model}
Upon extracting the seven obfuscation features, we customize the standard Z-score~\cite{curtis2016mystery} and compute an \emph{Obfuscation Z-score} as a quantitative metric for the degree of obfuscation applied to transfer operations. In particular, we calculate the cumulative distance between each feature's standard deviation and mean, and further compute the sum of standardized features to obtain the obfuscation Z-score for each contract:

\vspace{-0.2in}
\begin{equation}
Z_{\text{score}}=\sum_{i=1}^7\frac{x_i-\mu_i}{\sigma_i},
\end{equation}

where \(x_i\) is the \(i\)-th feature value for a given contract, and \(\mu_i\) and \(\sigma_i\) are the mean and standard deviation of the corresponding feature across the entire sample set. This Z-score represents the cumulative distance between the values of all seven features and their means, expressed in standard deviations, indicating the degree of obfuscation applied to the fund transfer operations.

\vspace{4pt}\noindent \textbf{Why Z-score.} Our seven features reside on heterogeneous scales (e.g., counts, ratios, graph-similarity scores), so summing raw values would be disproportionately influenced by the feature with the largest numeric range. Standardizing each feature using its corpus-level mean and standard deviation renders them unitless and comparable, ensuring that the aggregate score reflects joint extremeness rather than scale artifacts. The resulting Z-score quantifies how many standard deviations a contract’s combined behavior deviates from the corpus mean in the current snapshot, enabling corpus-agnostic tail selection (e.g., “top three-sigma”). Operationally, re-estimating $(\mu_i,\sigma_i)$ for each snapshot allows the score to adapt to ecosystem or compiler drift without manual, per-feature threshold tuning. In sum, Z-score normalization promotes fairness across features, interpretability across datasets, and stability over time.

\subsection{Evaluation of \codename}
We evaluate the performance of \codename on a dataset of 453 Ponzi-scam contracts, focusing on its effectiveness. Table~\ref{tab:dataset_classification} summarizes the data sources and our manual classification results.

\begin{table*}[t]
\centering
\small
\begin{minipage}[t]{.48\linewidth}
  \centering
\caption{Real-world Ponzi Scam Dataset}
\label{tab:dataset_classification}
\begin{tabular}{@{} l l r @{}}
\toprule
\textbf{Category} & \textbf{Item} & \textbf{Count} \\
\midrule
\multirow{2}{*}{Data source}
  & CRGB~\cite{Liang2024}   & 137 \\
  & SourceP~\cite{Lu2024}   & 316 \\
\cmidrule(lr){2-3}
\multicolumn{2}{r}{\textbf{Total}} & \textbf{453} \\
\midrule
\multirow{2}{*}{Classification}
  & Obfuscated      & 92 \\
  & Non-Obfuscated  & 361 \\
\bottomrule
\end{tabular}
\end{minipage}\hfill
\begin{minipage}[t]{.48\linewidth}
  \centering
  \small
\caption{Z-score statistics on the labeled dataset.}
\label{tab:zscore_statistics}
\begin{tabular}{@{} l r r r r r @{}}
\toprule
\textbf{Obfuscation} & \textbf{Count} & \textbf{Mean} & \textbf{Std} & \textbf{Min} & \textbf{Max} \\
\midrule
Non-Obfuscated & 361 & 4.571 & 0.641 & 2.581 & 5.261 \\
Obfuscated     &  92 & 6.888 & 3.587 & 4.688 & 26.456 \\
\addlinespace[2pt]
\midrule
\multicolumn{6}{@{}l}{\textbf{Welch's t-test:} $t=-6.172$, $p<10^{-6}$} \\
\bottomrule
\end{tabular}
\end{minipage}
\end{table*}

\begin{figure}[t]
    \centering
    \includegraphics[width=0.8\linewidth]{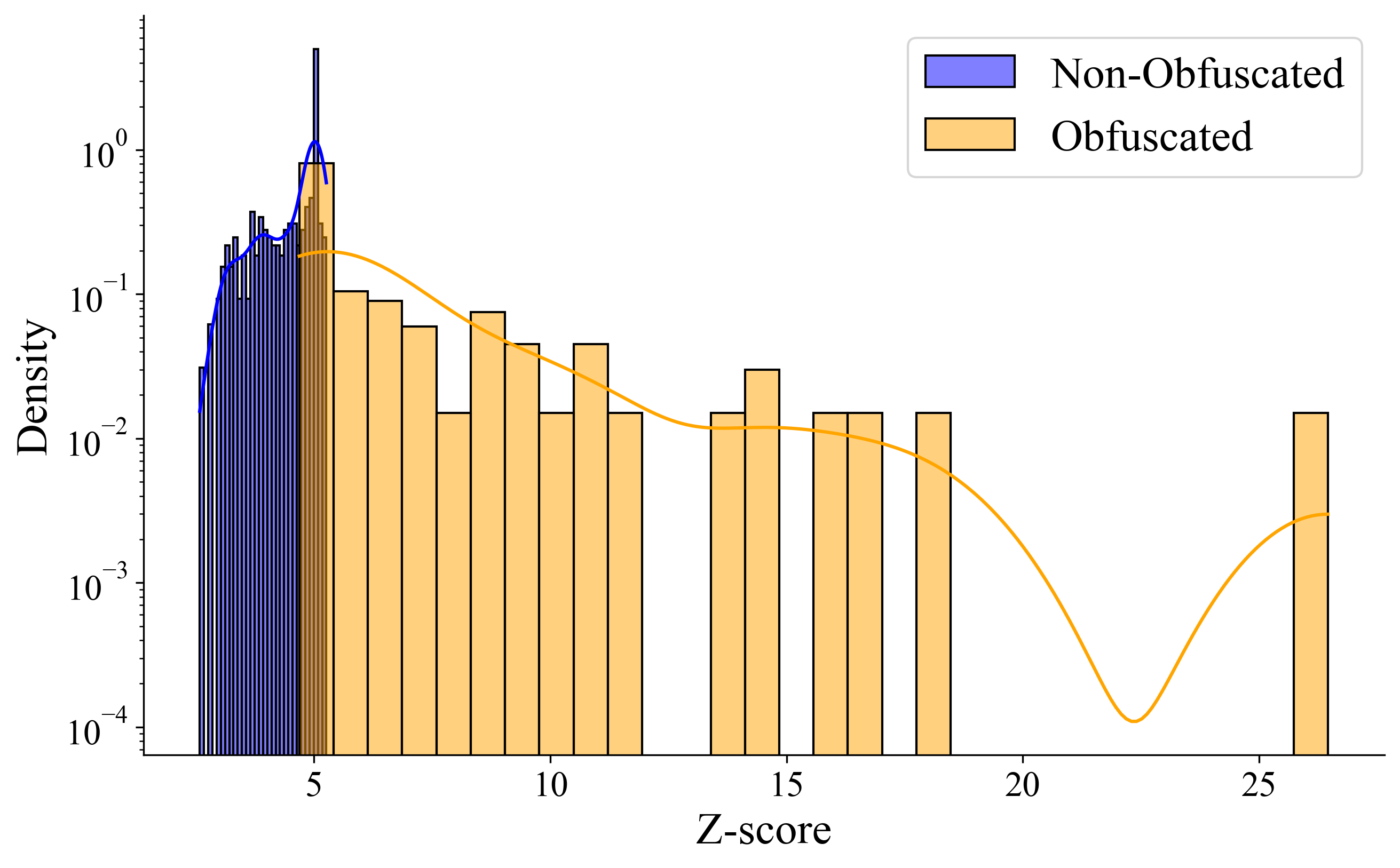}
    \vspace{-0.1in}
    \caption{Z-score distribution on Ponzi scam dataset.}
    \vspace{-0.15in}
    \label{fig:zscore_distribution}
\end{figure}

To evaluate the effectiveness of \codename, we compile the 453 contracts to obtain their bytecode and apply \codename to obtain the values of the seven obfuscation features, as well as the Z-scores. We manually check the extracted F1--F7 signals on these 453 contracts and find no extraction errors (feature-correctness check, $N=453$), which is different from detection accuracy. Furthermore, we examine the distribution of the Z-scores for the two groups (obfuscated vs. non-obfuscated), which is presented in Figure~\ref{fig:zscore_distribution}. As shown, the mean Z-score of the obfuscated group is significantly higher than that of the non-obfuscated group, and its standard deviation is also larger. This indicates that obfuscated contracts exhibit greater diversity and concealment. Additionally, we conduct a more detailed statistical investigation, as shown in Table~\ref{tab:zscore_statistics}. The results show that the difference between the Z-score distributions of the two groups is statistically significant (t = -6.172, \( p \approx 0 \)), confirming that \codename can effectively distinguish obfuscated contracts from non-obfuscated contracts by using Z-scores.

Table~\ref{tab:ponzi_metrics} summarizes standard detection metrics (i.e., precision, recall, ROC-AUC, and PR-AUC) for the proposed Z-score rule and a logistic-regression baseline trained on features F1–F7. Precision is the proportion of contracts predicted as obfuscated that are truly obfuscated, whereas recall is the proportion of obfuscated contracts correctly identified. The Z-score cutoff achieves perfect recall (1.0000) but low precision (0.2921), indicating a high false-positive rate; in contrast, logistic regression attains substantially higher precision (0.6130) at the cost of lower recall, reflecting the expected false-positive/false-negative trade-off on the same labeled set. Because the labels are imbalanced (92 obfuscated vs.\ 361 non-obfuscated), PR-AUC is more informative than accuracy, and moderate F1 scores are therefore expected in this setting. For the Z-score method, ROC-AUC and PR-AUC are computed by treating the raw Z-scores as continuous confidence values, while precision/recall/F1 are reported at the fixed threshold $Z \ge 4.637$. This threshold is derived from the labeled non-obfuscated subset of the Ponzi dataset as the upper bound of the 95\% confidence interval and then applied uniformly. Importantly, it is not tuned to optimize evaluation metrics. All logistic regression results are computed from out-of-fold predictions obtained under 5-fold cross-validation.

To assess which signals contribute most to predictive performance, we conduct a drop-column ablation: we remove one feature at a time, retrain the classifier, and re-evaluate. Table~\ref{tab:dropcolumn_importance} reports the three features whose removal yields the largest average decreases in PR-AUC and F1, where $\Delta$ values are averaged over the same 5-fold cross-validation and features are ranked by $\Delta$PR-AUC. Larger decreases indicate greater reliance on the removed feature; under this analysis, F5, F3, and F2 account for the largest performance degradations. This finding is consistent with the prevalence and tail analyses in Section~\ref{result}, that is, F3 is strongly enriched in the extreme tail, while F2 and F5 capture address manipulation and camouflage patterns that are particularly characteristic of Ponzi contracts.

In summary, the evaluation results show that \codename can accurately extract the predefined obfuscation features. In addition, it also shows that our Z-score representation model can effectively differentiate obfuscated from non-obfuscated smart contracts.

\begin{table*}[t]
\centering
\small
\caption{Detection metrics on the Ponzi dataset.}
\label{tab:ponzi_metrics}
\begingroup
\begin{tabular}{@{} l r r r r r @{} }
\toprule
\textbf{Method} & \textbf{Precision} & \textbf{Recall} & \textbf{F1} & \textbf{ROC-AUC} & \textbf{PR-AUC} \\
\midrule
Z-score cutoff ($\geq$ 4.637) & 0.2921 & 1.0000 & 0.4521 & 0.8463 & 0.6627 \\
LogReg (CV=5, 7 signals)      & 0.6130 & 0.4550 & 0.5186 & 0.7683 & 0.5888 \\
\bottomrule
\end{tabular}
\endgroup
\end{table*}

\begin{table}[t]
\centering
\small
\caption{Drop-column feature importance (Top-3).}
\label{tab:dropcolumn_importance}
\begingroup
\begin{tabular}{@{} l r r @{} }
\toprule
\textbf{Feature} & \textbf{$\Delta$ PR-AUC} & \textbf{$\Delta$ F1} \\
\midrule
F5 & 0.0744 & 0.1028 \\
F3 & 0.0261 & 0.0291 \\
F2 & 0.0222 & 0.0305 \\
\bottomrule
\end{tabular}
\endgroup
\end{table}

\begin{figure}[t]
    \centering
    \includegraphics[width=0.8\linewidth]{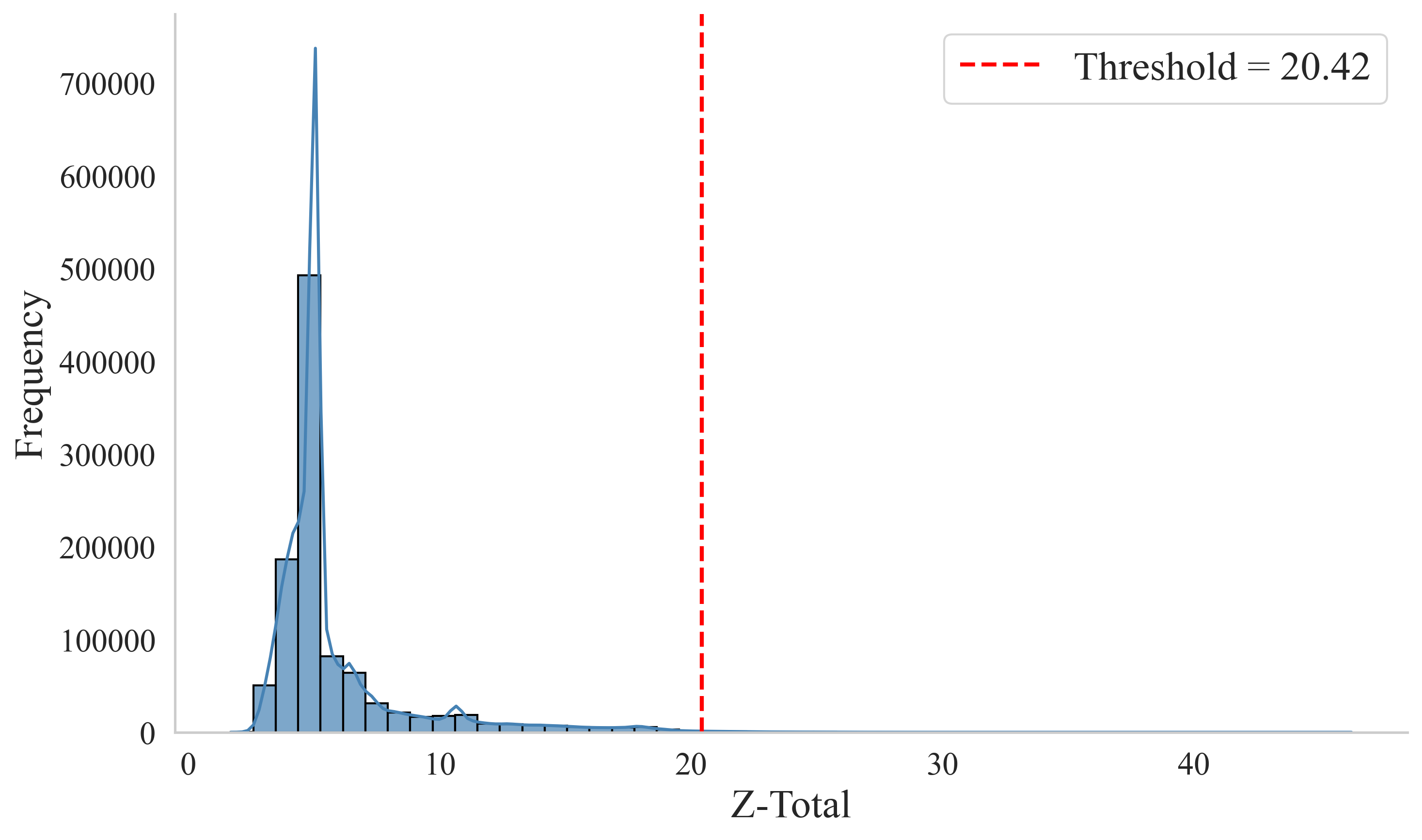}
    \vspace{-0.1in}
    \caption{Z-score distribution on mainnet (red dashed = top 3000 cutoff).}
    \vspace{-0.15in}
    \label{fig:zscore_dist_main}
\end{figure}

\vspace{4pt}\noindent \textbf{Static Analysis Efficiency.} The first pass yields SSA outputs for 1,042,923 contracts (93.5474\% first-run success) under a 20s timeout. The first-pass runtime averages 19.66s with a median of 19.9s. Failed contracts are diagnosed and rerun with the timeout removed; recovered outputs are merged back, so the prevalence analysis includes these rerun samples, indicating a stable pipeline. In the uncapped rerun, the average runtime is about 45s (median 41s), and large/complex bytecodes often fall in the 20--80s range, so the runtime is dominated by bytecode size and complexity.

\section{Smart Contract Obfuscations in the Wild}
\label{result}
In this section, we aim to answer RQ2 by studying the prevalence of obfuscation techniques in real world smart contracts. 

\subsection{Prevalence Study}

We collect 1,042,923 unique smart contract bytecodes that were \emph{active} on Ethereum mainnet between Jun 2022 and Oct 2024 (i.e., at least one on-chain interaction in this window), \emph{regardless of their original deployment dates}.

\vspace{4pt}\noindent \textbf{Distribution of Z-Score.} Figure~\ref{fig:zscore_dist_main} shows that the distribution of the obfuscation score (\texttt{Z-score}\ is strongly right‐skewed. Most contracts fall in the range \(3 \lesssim Z \lesssim 8\), with a peak near \(Z \approx 6\). For \(Z > 10\), the bar heights decline steadily, forming a long tail that extends to the highest scores. This visual pattern indicates that while most smart contracts exhibit a moderate level of obfuscation, a small subset employs an extremely high level. In Table~\ref{tab:zscore_stats_updated}, we summarize the distribution of Z-score across the 1,042,923 contracts. To examine the prevalence of obfuscated contracts, we use a Z-score threshold of 4.637 derived from real-world data in Table~\ref{tab:dataset_classification} to represent the 95\% confidence interval (CI) upper bound, calculated as 
\begin{equation}
  \vspace{-0.07in}
    4.571 \;+\; t_{0.975,360}\times\frac{0.641}{\sqrt{361}}
\;\approx\;4.637.
\end{equation}
Applying this cutoff, Table~\ref{tab:prevalence} reveals that 70.93\% of on-chain contracts deploy obfuscation at or above what would be considered a “normal” level. More than two-thirds of deployed contracts exceed the non-obfuscated CI ceiling, indicating that obfuscation has become routine in smart contract development. This pervasive adoption underscores the need for more robust analysis tools and transparency mechanisms to manage obfuscation in the Ethereum ecosystem. 

\begin{table*}[t]
\centering
\small
\begin{minipage}[t]{.48\linewidth}
  \centering
  \caption{Z-score statistics.}
  \label{tab:zscore_stats_updated}
  \begin{tabular}{@{} l r @{}}
    \toprule
    \textbf{Metric} & \textbf{Value} \\
    \midrule
    Count (contracts) & 1{,}042{,}923 \\
    Mean              & 5.867 \\
    Std               & 2.910 \\
    Min               & 1.698 \\
    Max               & 46.264 \\
    \bottomrule
\end{tabular}
\end{minipage}\hfill
\begin{minipage}[t]{.48\linewidth}
  \centering
  \caption{Prevalence of obfuscated smart contracts.}
  \label{tab:prevalence}
  \begin{tabular}{@{} l r r @{}}
    \toprule
    \textbf{Category} & \textbf{Count} & \textbf{Percentage} \\
    \midrule
    Above threshold ($>4.637$) & 739{,}763 & 70.93\% \\
    Below threshold ($<4.637$) & 303{,}160 & 29.07\% \\
    \bottomrule
\end{tabular}
\end{minipage}
\end{table*}

These results highlight that the key distinction in practice is degree rather than mere presence: most contracts cluster in a moderate range, whereas the most atypical behavior concentrates in the extreme right tail. The fact that 70.93\% of contracts exceed the non-obfuscated ceiling suggests that moderate obfuscation is commonplace, motivating prioritization of manual analysis on the top-3000 contracts with the highest Z-scores. This corpus-level prevalence provides a baseline for the remainder of our study: we focus on the extreme tail, where rare signals are concentrated, and auditing effort is expected to yield the highest return.

\vspace{4pt}\noindent \textbf{Prevalence of individual features.} Table~\ref{tab:feature_prevalence} reports, for each feature, its non-zero rate $P(F_i>0)$ and strength quantiles (median/P90/P99) over 1,042,923 contracts. F1, F4, and F5 occur in more than 70\% of contracts, indicating that multi-step address derivation, control-flow complexity, and camouflage instruction ratio are widespread in real-world bytecode. Consequently, feature presence alone is not a reliable risk indicator and should be interpreted jointly with magnitude (e.g., P90/P99). The distributions are markedly heavy-tailed (e.g., F1 increases from a median of 4 to a P90 of 478), showing that extreme contracts amplify these primitives by orders of magnitude. By contrast, F3 appears in only 8.50\% of contracts, suggesting that external-call obfuscation is less prevalent but potentially more discriminative when it occurs.

\begin{table}[t]
\centering
\small
\caption{Feature prevalence in 1,042,923 contracts.}
\label{tab:feature_prevalence}
\begingroup
\begin{tabular}{@{} l r r r r @{} }
\toprule
\textbf{Feature} & \textbf{P(F>0)} & \textbf{Median} & \textbf{P90} & \textbf{P99} \\
\midrule
F1  & 72.91\% & 4.00  & 478.00 & 974.00 \\
F2  & 41.20\% & 0.00  & 120.00 & 314.00 \\
F3  & 8.50\%  & 0.00  & 0.00   & 1.00   \\
F4  & 73.67\% & 4.00  & 25.00  & 43.00  \\
F5  & 73.82\% & 0.14  & 0.56   & 0.87   \\
F6  & 58.79\% & 56.03 & 98.77  & 99.90  \\
F7  & 39.79\% & 0.00  & 1.00   & 1.00   \\
\bottomrule
\end{tabular}
\endgroup
\end{table}

\vspace{4pt}\noindent \textbf{Prevalence of obfuscation patterns.} Using the T1--T7 $\leftrightarrow$ F1--F7 mapping, Table~\ref{tab:pattern_prevalence} reports pattern prevalence across the full corpus. T1, T4, and T5 are the most common (each $\approx$73\%), indicating that these primitives constitute a broad baseline of real-world obfuscation. T2 and T6 are moderately prevalent (41.20\% and 58.79\%), whereas T3 is comparatively rare (8.50\%), consistent with external-call obfuscation being less frequently deployed. As high-frequency base patterns, T1/T4/T5 are not, by themselves, strong indicators of maliciousness; in contrast, the rarer T3 pattern can provide a sharper signal in prioritization settings.

\begin{table}[t]
\centering
\small
\caption{Pattern prevalence in 1,042,923 contracts.}
\label{tab:pattern_prevalence}
\begingroup
\begin{tabular}{@{} l r @{} }
\toprule
\textbf{Pattern} & \textbf{P\_all} \\
\midrule
T1 & 72.91\% \\
T2 & 41.20\% \\
T3 & 8.50\%  \\
T4 & 73.67\% \\
T5 & 73.82\% \\
T6 & 58.79\% \\
T7 & 39.79\% \\
\bottomrule
\end{tabular}
\endgroup
\end{table}

\vspace{4pt}\noindent \textbf{Decomposition of Z-score.} Table~\ref{tab:zscore_contrib} decomposes the overall Z-score into feature-level shares, computed as mean absolute standardized contributions. F3 accounts for the largest share (18.04\%), followed by F7 (16.69\%) and F6 (16.33\%), indicating that these components contribute disproportionately to the aggregate score. F2 contributes the smallest share (10.90\%). Overall, high Z-scores are driven primarily by strong, less ubiquitous signals rather than by the most common background patterns.
\begin{table}[t]
\centering
\small
\caption{Z-score contribution share across the corpus.}
\label{tab:zscore_contrib}
\begingroup
\begin{tabular}{@{} l r @{} }
\toprule
\textbf{Feature} & \textbf{Share\_all} \\
\midrule
F1  & 12.03\% \\
F2  & 10.90\% \\
F3  & 18.04\% \\
F4  & 13.13\% \\
F5  & 12.88\% \\
F6  & 16.33\% \\
F7  & 16.69\% \\
\bottomrule
\end{tabular}
\endgroup
\end{table}

\begin{figure*}[t]
    \centering
    \includegraphics[width=0.95\textwidth]{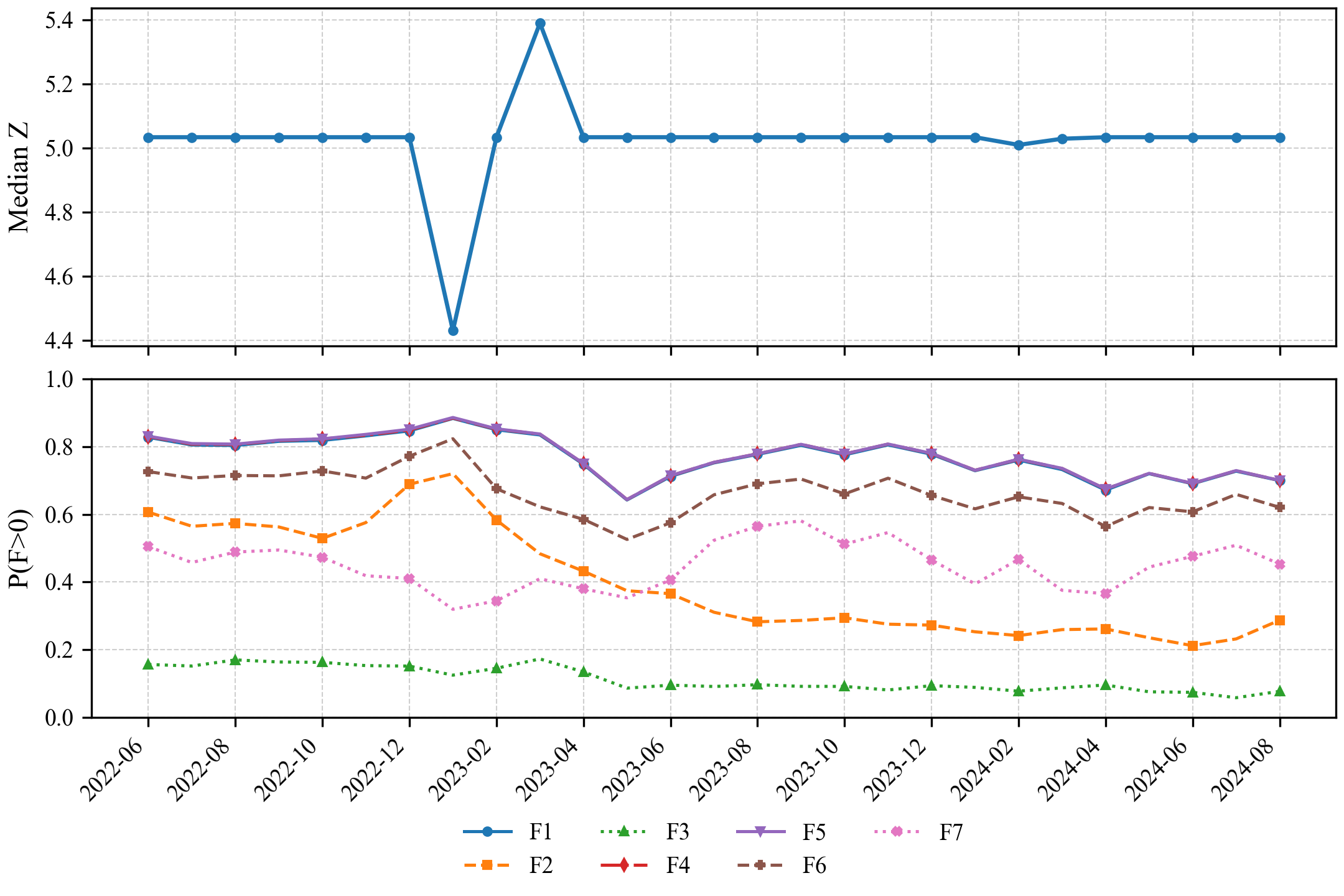}
    \caption{Temporal trend of obfuscation signals (2022--2024): monthly median Z-score (top) and P(Fi>0) for all features (bottom).}
    \label{fig:trend_monthly}
\end{figure*}

\vspace{4pt}\noindent \textbf{Temporal trend (2022--2024).} To quantify the trend for RQ2, we associate each contract with its creation timestamp and compute monthly aggregates from June 2022 through August 2024 (UTC). Figure~\ref{fig:trend_monthly} plots the monthly median Z-score (top) and the prevalence of each feature (bottom). The median Z-score remains largely stable around $\sim 5.0,$ while individual feature prevalences vary within bounded ranges, suggesting a stable aggregate baseline with shifts in the relative emphasis of specific techniques. Notably, the prevalence curves for F1/F4/F5 nearly overlap (differences $<0.5$ percentage points), consistent with these high-frequency primitives co-occurring as baseline obfuscation; consequently, their month-to-month movements are tightly coupled. Our dataset includes contract creations up to 2024-08; subsequent months contain no new deployments in this snapshot. These observations motivate the extreme-tail analysis in Section~\ref{top0.3}.

\rqanswer{2}{Applying \codename to real-world Ethereum smart contracts reveals that most real-world smart contracts exhibit a moderate level obfuscation, and there
exists a small subset employing an extreme higher level of obfuscations; 
}

\subsection{Analysis of the 3000 highly-obfuscated Contracts}
\label{top0.3}
To better understand the impact of obfuscation techniques used in real-world smart contracts, we select the top 3000 contracts in terms of Z-scores as highly suspicious targets for in-depth analysis. We focus our analysis on top 3000 contracts based on two considerations. First, in statistics, the rightmost 3000 samples (roughly 0.3\% of the total) in a normal distribution typically indicate extreme outliers~\cite{walfish2006review,cousineau2010outliers} (i.e., values exceeding approximately three standard deviations from the mean). Second, focusing on the top 3,000 samples can significantly reduce our manual efforts while still adequately covering the typical values of the distribution. We hence limit our \emph{manual analysis and case studies} to the top 3,000 contracts to strike a balance between feasibility and coverage. 

\vspace{4pt} \noindent\textbf{Overview of the top 3000 contracts:} Table~\ref{tab:top3k_compact} presents an overview of the top 3000 contracts. It can be seen that the top 3000 contracts' Z-score is \emph{heavy-tailed}, evidenced by a median of 21.922, a minimum of 20.419, and a maximum of 46.264. In addition, all 3000 contracts (100\%) have exhibited at least 2 obfuscation features, and 98.1\% have exhibited at least 3 obfuscation features. Among the 3000 contracts, 46.0\% have their source code published on Etherscan. We distinguish explicit obfuscation (verified source code visible to users) from covert obfuscation (bytecode-only or unverified). In the top-3000 tail, only 46.0\% are explicit by this definition, leaving the majority covert. This coupling between deep obfuscation and code invisibility raises audit cost and motivates focusing manual efforts on the extreme tail. Table~\ref{tab:top3k_bins} shows the distribution of the top 3000 contracts' Z-score. It can be seen that the majority of the contracts' Z-score fall into the range of 20–25 (88.1\%). 11.1\% of them fall into the range of 25 - 30. Less than 1\% of them fall into the range of 30 - 50.
In short, the top-3000 set is the actionable slice for auditing: stacked obfuscation and low source visibility make it the highest-yield target for further investigation.

\begin{table*}[t]
\centering
\parbox[t]{.48\linewidth}{
  \centering
  \caption{Top 3000 obfuscated contracts.}
  \label{tab:top3k_compact}
  \begin{tabular}{@{}lc@{}}
    \toprule
    \textbf{Metric} & \textbf{Value (Share)} \\
    \midrule
    $z_{\text{score}}$ min / median / max & 20.419 / 21.922 / 46.264 \\
    features $\ge 2$ & 3{,}000 (100.0\%) \\
    features $\ge 3$ & 2{,}943 (98.1\%) \\
    With verified source code & 1{,}380 (46.0\%) \\
    \bottomrule
  \end{tabular}
}
\hfill
\parbox[t]{.48\linewidth}{
  \centering
  \caption{Distribution of $z_{\text{score}}$ in top 3000 contracts.}
  \label{tab:top3k_bins}
  \begin{tabular}{@{}lc@{}}
    \toprule
    \textbf{$z_{\text{score}}$ range} & \textbf{Count (Share)} \\
    \midrule
    20--25 & 2{,}644 (88.1\%) \\
    25--30 & 335 (11.1\%) \\
    30--35 & 16 (0.5\%) \\
    35--40 & 2 (0.1\%) \\
    40--45 & 2 (0.1\%) \\
    45--50 & 1 (0.0\%) \\
    \bottomrule
  \end{tabular}
}
\end{table*}

\vspace{4pt}\noindent \textbf{Top-3000 feature prevalence.} Table~\ref{tab:top3k_feature_prevalence} shows that the extreme tail is dominated by F1/F2/F4/F5/F6 (near-universal presence), while F3 exhibits strong enrichment (11.44x) and F7 remains present in 72.87\% of top-3000 contracts. The near-universal coverage indicates that the extreme tail reflects stacked obfuscation rather than a single trick, and F3's enrichment makes external-call hiding a key gate into the tail.

\begin{table}[t]
\centering
\small
\caption{Top-3000 feature prevalence and enrichment.}
\label{tab:top3k_feature_prevalence}
\begingroup
\begin{tabular}{@{} l r r r @{} }
\toprule
\textbf{Feature} & \textbf{P\_top3k} & \textbf{Enrichment} & \textbf{Median\_top3k} \\
\midrule
F1  & 100.00\% & 1.37x & 1199.00 \\
F2  & 100.00\% & 2.43x & 417.00  \\
F3  & 97.23\%  & 11.44x & 1.00   \\
F4  & 100.00\% & 1.36x & 33.00  \\
F5  & 100.00\% & 1.35x & 0.11   \\
F6  & 99.13\%  & 1.69x & 98.40  \\
F7  & 72.87\%  & 1.83x & 1.00   \\
\bottomrule
\end{tabular}
\endgroup
\end{table}

\vspace{4pt}\noindent \textbf{Ablation on top-3000 stability.} Removing F3 yields the largest overlap drop (40.43\%), followed by F2 (27.37\%) and F4 (19.37\%), indicating these signals most strongly shape the extreme tail ranking. By contrast, F5/F6/F7 have much smaller drops; they act as background noise rather than decisive ranking drivers.

\begin{table}[t]
\centering
\small
\caption{Top-3000 ranking stability under feature ablation.}
\label{tab:top3k_ablation}
\begingroup
\begin{tabular}{@{} l r r @{} }
\toprule
\textbf{Feature} & \textbf{Overlap@3000} & \textbf{Drop} \\
\midrule
F1  & 81.20\% & 18.80\% \\
F2  & 72.63\% & 27.37\% \\
F3  & 59.57\% & 40.43\% \\
F4  & 80.63\% & 19.37\% \\
F5  & 96.23\% & 3.77\%  \\
F6  & 97.70\% & 2.30\%  \\
F7  & 96.33\% & 3.67\%  \\
\bottomrule
\end{tabular}
\endgroup
\end{table}

\vspace{4pt}\noindent \textbf{TopK sensitivity.} The enrichment trends remain stable across Top-1000/3000 (Table~\ref{tab:topk_sensitivity}), indicating the extreme tail is a structural phenomenon rather than a threshold artifact.

\begin{table}[t]
\centering
\small
\caption{TopK sensitivity for selected features.}
\label{tab:topk_sensitivity}
\begingroup
\begin{tabular}{@{} r r r @{} }
\toprule
\textbf{K} & \textbf{F3} & \textbf{F7} \\
\midrule
1000 & 98.30\% & 72.00\% \\
3000 & 97.23\% & 72.87\% \\
\bottomrule
\end{tabular}
\endgroup
\end{table}

\vspace{4pt} \noindent\textbf{A close look at the top 3000 contracts.} To facilitate further analysis, we examine the top 3000 contracts ranked by Z-score and use total funds to prioritize risk assessment during manual inspection. Our manual investigation reveals that all contracts employ obfuscation techniques, with 463 contracts posing very \emph{high risks} due to extensive obfuscation that conceals four potentially malicious and suspicious behaviors. Such 463 contracts can be divided into four categories: MEV bots, Ponzi schemes, fake decentralization, and extreme centralization. The total amount of funds absorbed by these contracts reaches approximately \$100 million USD in Ether. Table~\ref{tab:case_coverage_top3k} summarizes the distribution of the 463 contracts among the four categories. In general, MEV bots account for the largest share (50.1\%), followed by extreme centralization (33.0\%). Ponzi and Fake Decentralization are less frequent but salient for risk analysis. Below, we conduct a detailed case study of the four contract types, discussing their typical obfuscation patterns, associated financial impacts, and the lifespan of their transaction activity.

\begin{table}[t]
  \centering
  \caption{Distribution of the contracts in the four case studies.}
  \label{tab:case_coverage_top3k}

  \begin{tabular}{@{}lrr@{}}
    \toprule
    \textbf{Case-study pattern} & \textbf{Count} & \textbf{Percentage} \\
    \midrule
    MEV bots & 232 & 50.1\% \\
    Extreme Centralization & 153 & 33.0\% \\
    Ponzi & 48 & 10.4\% \\
    Fake Decentralization (renounce) & 30 & 6.5\% \\
    \midrule
    Total instances (multi-label) & 463 & 100\% \\
    \bottomrule
  \end{tabular}%
  \vspace{-0.15in}
\end{table}

\subsection{Case I: MEV Bots}
\label{subsec:mev-bots}

Due to varying exchange rates across multiple decentralized exchanges (DEXs), arbitrage opportunities arise. MEV (Maximal Extractable Value, formerly Miner Extractable Value) refers to the value extracted by ordering, inserting, or censoring transactions within a block. MEV itself is not inherently malicious. In this work, we focus on bot behaviors whose obfuscated transfer paths and fund flows raise security concerns. MEV bot contracts are thus developed and deployed to exploit the opportunities and gain profits with techniques such as front-running~\cite{daian2019flash}, back-running~\cite{qin2022quantifying}, and sandwich attacks~\cite{daian2019flash,mazorra2022price}. Our study finds that some MEV bot contracts leverage \emph{heavily obfuscated} to hide their profit-making logic and thwart analysis. Our analysis of the highly obfuscated MEV bot contracts reveals four unique MEV-specific obfuscation patterns that combine multiple obfuscation techniques from our taxonomy.

\begin{itemize}[leftmargin=*]
\item[(1)] \textbf{Fallback only.} In this pattern, MEV bot contracts only implement the fallback functions to parse the calldata, which then jump to the corresponding location to continue the execution. Eliminating the 4-byte function selectors introduces additional challenges for static analysis, including function identification and call graph generation. The code typically features numerous SWAP and JUMPI instructions. This strategy is highly relevant to F3 and F4 features in Table~\ref{tab:feature_summary} because the fallback function typically relies on external calls to handle \texttt{calldata} and uses conditional branches to increase the complexity. A representative MEV bot employing this obfuscation pattern can be found at address \hash{0x6b75d8af000000E20b7a7ddf000Ba900b4009A80}.

\item[(2)] \textbf{ABI distortion.} Some MEV bots manipulate function selectors by shortening or relocating them within \texttt{calldata}, making it difficult to identify entry points of the contract. This pattern is relevant to features F1, F2, and F4 because (1) it results in complex address generation processes, involving multiple steps; (2) manipulating ABI elements involves string operations like concatenation or hashing; and (3) it may introduce additional control flow complexity by introducing branches. A representative MEV bot employing this obfuscation pattern can be found at address \hash{0x1F2f10d1C40777AE1da742455c65828fF36df387}.

\item[(3)] \textbf{Address obfuscation.} This pattern uses operations like \texttt{PUSH4} and \texttt{XOR} to reconstruct the beneficiary address, and requires precisely-length \texttt{calldata} inputs, immediately reverting on mismatch. It is related to features F1 and F2 in that it introduces multiple steps to dynamically construct the transfer address and sometimes involves string manipulations. Moreover, our observation shows that it often introduces many irrelevant instructions, reducing the proportion of transfer-related instructions. Hence, it is directly captured by F1. A representative MEV bot employing this obfuscation pattern can be found at address \hash{0xA69babEF1Ca67a37fFAf7a485DfFf3382056E78c}.

\item[(4)] \textbf{Runtime constraints.} This pattern introduces conditional branches based on chain-specific variables (e.g., block.coinbase), preventing frontrunning in the public mempool by directing different logic flows depending on the block builder. We find that this strategy is related to features F4 and F7, as it introduces additional conditional branches in the control flow and often emits irrelevant or misleading logs. A representative MEV bot employing this obfuscation pattern can be found at address \hash{0x51C72848C68A965F66fA7A88855F9F7784502a7F}.
\end{itemize}

\begin{figure*}[htbp]
    \centering
    \includegraphics[width=0.7\linewidth]{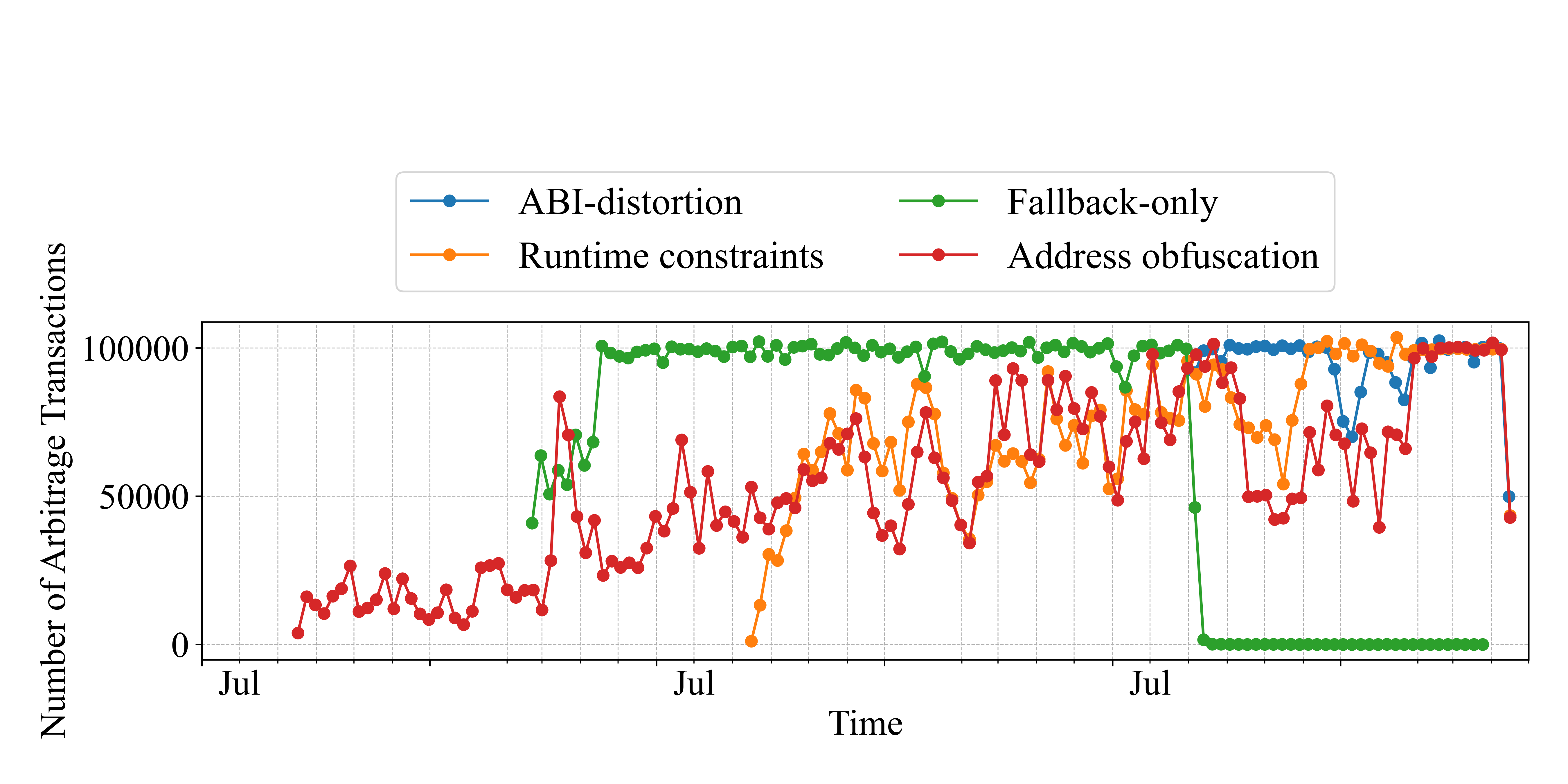}
    \vspace{-0.25in}
    \caption{Time Series Analysis of transaction Volumes}
    \vspace{-0.15in}
    \label{fig:mev-ts}
\end{figure*}

\vspace{4pt}\noindent \textbf{Arbitrage transactions of MEV bots.} To gain more insights into the impact of different obfuscation patterns adopted by MEV bots, we select the representative MEV bots from each pattern to analyze their transaction activities. We draw several observations from a time-series analysis of arbitrage transactions submitted to each MEV bot, as shown in Figure~\ref{fig:mev-ts}. Notably, the fallback‐only MEV bot shows a brief burst of high activity early in the period before dropping to near-zero. The ABI distortion contract shows more sustained volume, while runtime-constraint and address-obfuscation patterns exhibit greater variability in transaction counts.
These distinct profiles show that obfuscation pattern choice aligns with observable activity profiles, reinforcing the value of pattern-level analysis.

\subsection{Case II: Ponzi Schemes}
\label{Ponzi/Pyramid Scheme Contracts}

Ponzi contracts encourage users to deposit funds or purchase specific tokens by claiming high yield returns through automatic buybacks and burns, compounded mining, or cross-platform arbitrage. Then, they force participants to hold the tokens, implement multi-level commission systems, and rely on new funds from subsequent investors to support returns for earlier participants~\cite{gazi2024code}, exhibiting classic Ponzi characteristics. When additional funds are short, the project creator dumps tokens to reap enormous profits, triggering a system collapse and causing losses for participants~\cite{Chen2021,Liang2024}. We investigate obfuscated contracts and identify some representative Ponzi-specific behaviors (e.g., multi-level deduction, referral/downlines).

\noindent \textbf{Obfuscation patterns.} We exemplify how the contract obfuscates the logic of forcing token holding, implementing buyback and burn mechanisms, and using multi-layered function wrappers.

\begin{itemize}[leftmargin=*]
    \item[(1)] \textbf{Multi-level deduction and address generation.}  
    When users withdraw tokens, a cumbersome fee-deduction process is required, which involves parameters such as devTreasury, refBonus, and buyNBurn. \codename performs backward slicing from transfer operations and detects an obfuscated computational process. Furthermore, we see that the owner controls these parameter configurations and can adjust them at will.
    
    \item[(2)] \textbf{Layered Logic Based on External Inputs (Referral/Downlines).}  
    We identify a recruitment mechanism in the contract, which is a multi-level data structure. This indicates that user returns do not come from the contract’s own operations, but are instead distributed from the funds of new investors.
    
    \item[(3)] \textbf{Abundant "Buyback and Holding Check" Strings/Events.} 
    By analyzing event names or string constants, \codename detects terms like “Burn”, “MLMReward”, or “RetirementYeld”. These words are typically associated with forced token holding, token burning, or multi-level commissions, typical keywords of Ponzi/pyramid schemes.
    
\end{itemize}

\vspace{4pt}\noindent\textbf{A representative contract:}
Deployed at \hash{0x25cb947ebef1c56e14d5386a80262829739dbdbe} on the Ethereum mainnet, the contract exhibits the above obfuscation patterns. Over \emph{1,460} days (2020-07-29--2024-07-29), the contract involves \emph{27,303} interactions that all transferred ETH (with 0 ERC-20 token transfer), totaling 1,433.21 ETH. This contract is found matching the following obfuscation patterns:
(i) \textbf{Multi-level deduction.} The withdrawal path computes layered splits into \texttt{devTreasury}, \texttt{refBonus}, and \texttt{buyNBurn} sinks; these ratios are owner-configurable and can be adjusted at will, producing opaque fee chains.
(ii) \textbf{Referral/downlines.} The call data encodes a referrer and users are positioned in a matrix-like structure; payouts to earlier participants are funded by subsequent deposits rather than any productive on-chain activity.
(iii) \textbf{Buyback/holding narrative.} The code and logs contain terms such as ``Burn'', ``MLMReward'', and misspelled variants (e.g., ``RetirementYeld''), alongside frequent ``register/upgrade/reinvest/missed-earnings'' events and a payout structure that redistributes funds to uplines. Through this example, we show that Ponzi contracts typically leverage heavy obfuscation techniques to hide their business logic to avoid detection. To our best knowledge, our work is the first to report this contract as a Ponzi scheme.

\subsection{Case III: Fake Decentralization}
\label{subsubsec:fake-renounce}
This type of contract claims to decentralize control to attract participants, while maintaining obfuscated backdoor functions that allow owners to control the contract. Our study shows that two components are used to implement its malicious logic. The first one is called \texttt{fake renouncement of ownership}. Specifically, the contract states that executing \texttt{renounceOwnership()} can remove centralized ownership. However, the function merely transfers ownership to another address under the project owner's control, or does nothing, except emit a seemingly correct log to deceive participants. The second component is malicious backdoors, which are obfuscated functions (e.g., \texttt{Failsafe} or \texttt{Emergency}). They are claimed to handle system crashes, but in fact allow the project owner to withdraw assets at any time, resulting in financial losses for participants.

\vspace{4pt}\noindent \textbf{Obfuscation patterns.} Through our detection and investigation, we found that this type of contract adopts three obfuscation patterns. First, the contract duplicates ownership-transfer logic across multiple functions (feature F6), which differ only in parameter or variable names, increasing code complexity and hindering static analysis and manual auditing. Second, the contract often hides its core logic by inserting many empty and useless code segments (feature F5), making it difficult for auditors to quickly pinpoint the core backdoor. Third, the contract publicly claims that “control has been relinquished,” while the \texttt{onlyOwner} modifier remains effective. Alternatively, it may contain a function (e.g., \texttt{\_transferOwnership(addr)}), where \texttt{addr} is an address controlled by the owner, then the actual control is still maintained. 

\vspace{4pt}\noindent\textbf{A representative contract:}
The above patterns are instantiated in a contract deployed at \hash{0xb194A96AADC7e99a2462EF1669eB38E6B541DF79}. It involves \emph{11} transactions over \emph{3} days (2020-10-06 to 2020-10-10). During this period, it receives $0$ ETH and handles one ERC-20 token, \emph{YELD}, with $5{,}000$ in and $5{,}000$ out (gross $10{,}000$, net $0$).
The contract is found matching the following patterns:
(i) {Fake renouncement}: the contract claims ownership renouncement while preserving effective owner control via aliased transfer/renounce paths;
(ii) {Misleading logs}: the contract emits “relinquished control” log events. However, the \texttt{onlyOwner} modifier still remains effective;
(iii) {Code padding/noise}: the contract uses redundant stubs and near-duplicate permission paths without changing semantics.

\subsection{Case IV: Extreme Centralization}
\label{subsec:extremely-centralized}
Contracts in this category share a common design principle: all critical operations, from permission management to fund extraction, are ultimately controlled by a single address, despite superficial “decentralized” interfaces or multi-role declarations. Three typical manifestations are:

\begin{itemize}[leftmargin=*]
  \item \textbf{Centralized permission control:} Roles such as \texttt{admin}, \texttt{liquidateAdmin} (or \texttt{manager}, \texttt{\_super}, etc.) are all initialized to the deployer’s address, allowing unilateral modification of oracles, collateral ratios, fee structures, and forced liquidations.
  \item \textbf{Arbitrarily adjustable fee/tax:} They impose exorbitant buy/sell/withdrawal fees (often 10\%–50\%, up to 99\%) via an \texttt{onlyOwner}-protected function, with all collected fees routed to one address.
  \item \textbf{Lack of funding lock.} Some contracts, such as "staking" or "farming," claim to have emergency withdrawals. However, there are often functions (e.g., emergencyWithdraw(), requestWithdraw(), or claimTokens()) under the owner's control that can enable instant drainage.
\end{itemize}


\vspace{4pt}\noindent \textbf{Obfuscation patterns.} We investigate the typical obfuscation patterns adopted by these contracts and found that they often use redundant functions, events, and misleading names to obscure the centralized control, as revealed by features F6 and F7. Here, we list a few very representative ones.

\begin{itemize}[leftmargin=*]
  \item[(1)] \textbf{Role masquerading.} They tend to create multiple roles that point to the same address.
    \begin{lstlisting}[language=Solidity]
constructor() {
    admin = msg.sender;
    liquidateAdmin = msg.sender;
}
    \end{lstlisting}
    
  \item[(2)] \textbf{Redundant permission checks.} They introduce identical checks repetitively to inflate code complexity without adding any real safety. 
  \begin{lstlisting}[language=Solidity]
require (msg.sender == admin); 
require(msg.sender == liquidateAdmin);
    \end{lstlisting}
    
  \item[(3)] \textbf{Dynamic fee adjustment:} This function lets the owner unilaterally change both buy and sell tax rates at any time, enabling arbitrary fee hikes that can extract maximum revenue from users without prior notice.
    \begin{lstlisting}[language=Solidity]
function setTaxes(uint256 buyTax, uint256 sellTax) external onlyOwner {
    taxForBuy = buyTax;
    taxForSell = sellTax;
}
    \end{lstlisting}
    
  \item[(4)] \textbf{Backdoored withdrawals.} Although labeled as an emergency rescue, this owner-only method allows immediate token transfers from the contract to the owner’s EOA, effectively serving as a hidden backdoor to drain all funds.
    \begin{lstlisting}[language=Solidity]
function emergencyWithdraw(uint256 amount) external onlyOwner {
    token.safeTransfer(msg.sender, amount);
}
    \end{lstlisting}
    
  \item[(5)] \textbf{Redundant event and function:} 
    The contract may contain hundreds of emit calls (e.g., \ FeeEvent, UserUnlocked) and dozens of near-duplicate functions (e.g., \_swapTokens, \_addLiquidity, \_withdrawFromBank, fulfillDeposited) interleaved to generate noise for auditing.
\end{itemize}

\vspace{4pt}\noindent\textbf{A representative contract:} The above patterns are exemplified in a contract deployed at \hash{0xd7caa679aa6e39c3891bd7a63b058bb8a269da52}. Our analysis shows that it involves \emph{439} transfer transactions within \emph{81} days (2023-06-14 to 2023-09-03). The total received ETH is $0.2006$ ETH. The 
received ERC-20 tokens include 10 \emph{USDT} and $19.074240298$ \emph{APE}. The contract is found matching the following patterns:
(i) Role masquerading: multiple roles are ultimately pointing to the same address, resulting in centralized control;
(ii) {Backdoored withdrawals}: There are owner-controlled functions such as \texttt{withdraw}/\texttt{withdrawErc20}, which enable immediate extraction of funds in the contract.

\vspace{4pt}\noindent \textbf{Summary.} With the above four case studies, we show that it is prevalent for real-world contracts to employ various obfuscation techniques to hide their malicious behaviors such as MEV bots, Ponzi schemes, and fake decentralization, posing security risks to users.

\section{Financial Impact of Smart Contract Obfuscation}
\label{financial}
Following the detailed behavioral analysis, we further investigate the financial impact of obfuscation in the real world to answer RQ3. To achieve this, we collect a representative dataset of scam smart contracts, use \codename to detect obfuscation, and quantitatively compare the financial impact of obfuscated and non-obfuscated scam smart contracts to gain a deeper understanding of how obfuscation techniques affect scam financial gain.
Consistent with the extreme-tail focus in Section~\ref{result}, we interpret impact through tail behavior rather than average losses.

\subsection{Dataset}
\label{sec:financial-dataset}
We leverage the dataset from a prior study~\cite{Li2023}, which conducted a large-scale study of scams in the wild and reported approximately 13K scam contracts (which we refer to as the Li2023 dataset). To obtain ground truth on code obfuscation, we manually examine the dataset and label each contract into two groups: with-obfuscation and no-obfuscation, containing 9,197 and 3,826 unique contracts, respectively.

\subsection{Overview of Financial Loss}
After obtaining the two groups of contracts, we first analyze the inbound funds (fund inflows) of each group and the involved victims on the Ethereum mainnet to compare the financial loss (focus on ETH only) and then quantify the involved victims by counting unique Externally Owned Addresses (EOA) that directly send ETH to the contracts. Table~\ref{tab:inbound_eth_stats} shows the statistics of inbound funds of each contract group. It can be seen that the average inbound funds between the two groups are very close (0.3403 ETH vs.\ 0.3455 ETH), and the median values are also at the same order (0.06 vs.\ 0.10 ETH). However, the maximum inbound funds for obfuscated contracts (201.74 ETH) is about $2.41\times$ that of non-obfuscated contracts (83.62 ETH), indicating that obfuscated contracts are more likely to cause severe financial damage to victims. 

\begin{table}[h]
  \centering
  \caption{Inbound funds statistics.}
  \label{tab:inbound_eth_stats}
  \resizebox{0.7\columnwidth}{!}{%
  \begin{tabular}{l|c|cccc}
\hline
\multicolumn{1}{c|}{\multirow{2}{*}{\textbf{Contract Group}}} & \multirow{2}{*}{\textbf{Count}} & \multicolumn{4}{c}{\textbf{Inbound Funds (ETH)}}              \\ \cline{3-6} 
\multicolumn{1}{c|}{}                                         &                                 & \textbf{Sum} & \textbf{Median} & \textbf{Mean} & \textbf{Max} \\ \hline
No-Obfuscation                                                & \multicolumn{1}{r|}{3,826}      & 1,321.82     & 0.10            & 0.3455        & 83.62        \\
With-Obfuscation                                              & \multicolumn{1}{r|}{9,197}      & 3,129.997    & 0.06            & 0.3403        & 201.74       \\ \hline
\end{tabular}%
}
\end{table}

Although mean and median inbound funds are close, the 2.41x gap in maxima indicates that obfuscation amplifies tail losses rather than average losses, which is the more relevant signal for risk management.

\subsection{Timeline Trend of Financial Loss}
To better illustrate the financial impact of obfuscation, we also analyze the timeline trend of victims and the inbound funds caused by the two contract groups, from 2018 to 2025 in a 15-day interval.

\begin{figure}[t!]
    \centering
    \subfloat[Inbound funds aggregated by a 15-day interval.]{%
        \includegraphics[height=1.8in]{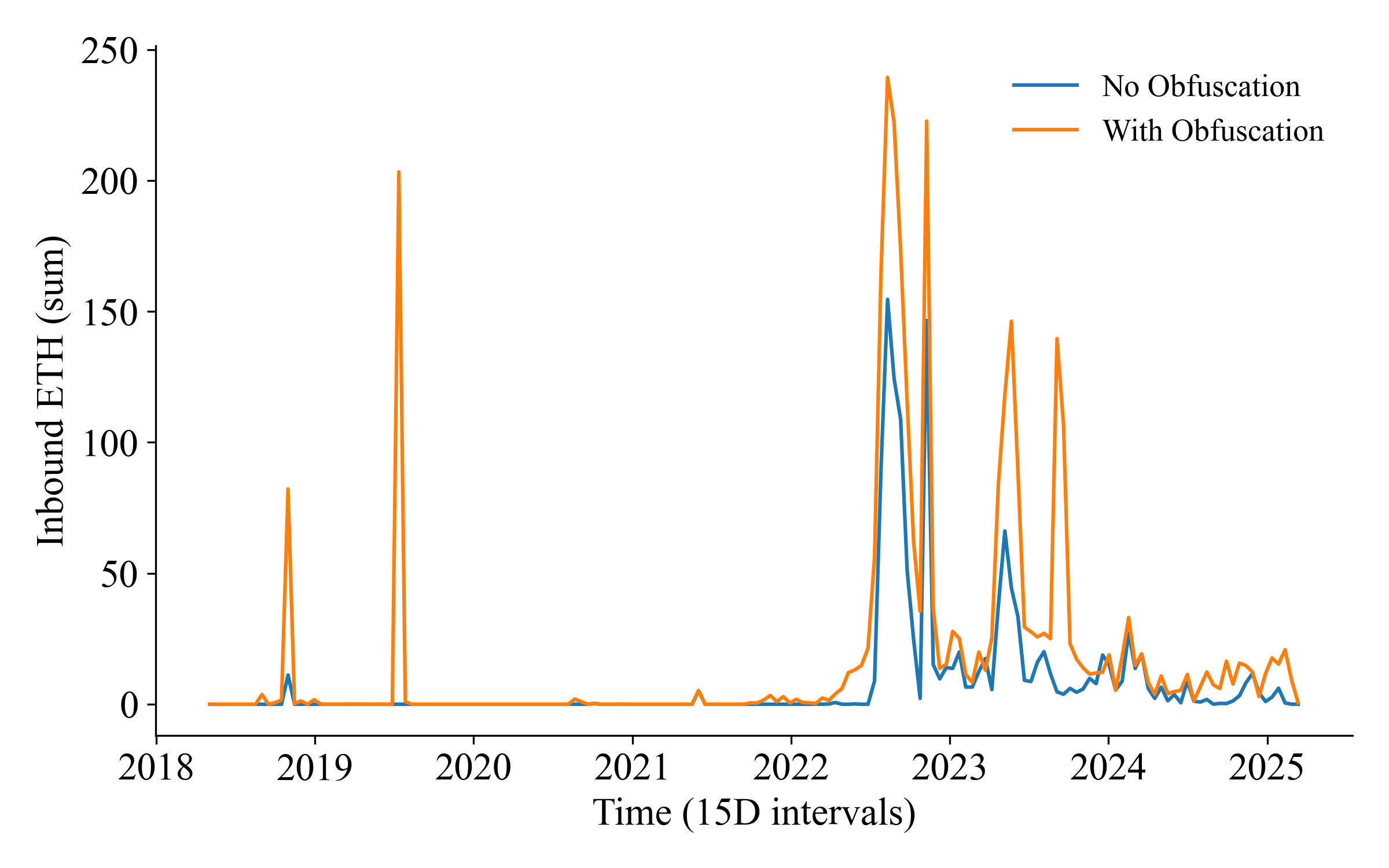}%
        \label{victim_inbound}%
    }\hfil
    \subfloat[Victim EOA aggregated by a 15-day interval.]{%
        \includegraphics[height=1.8in]{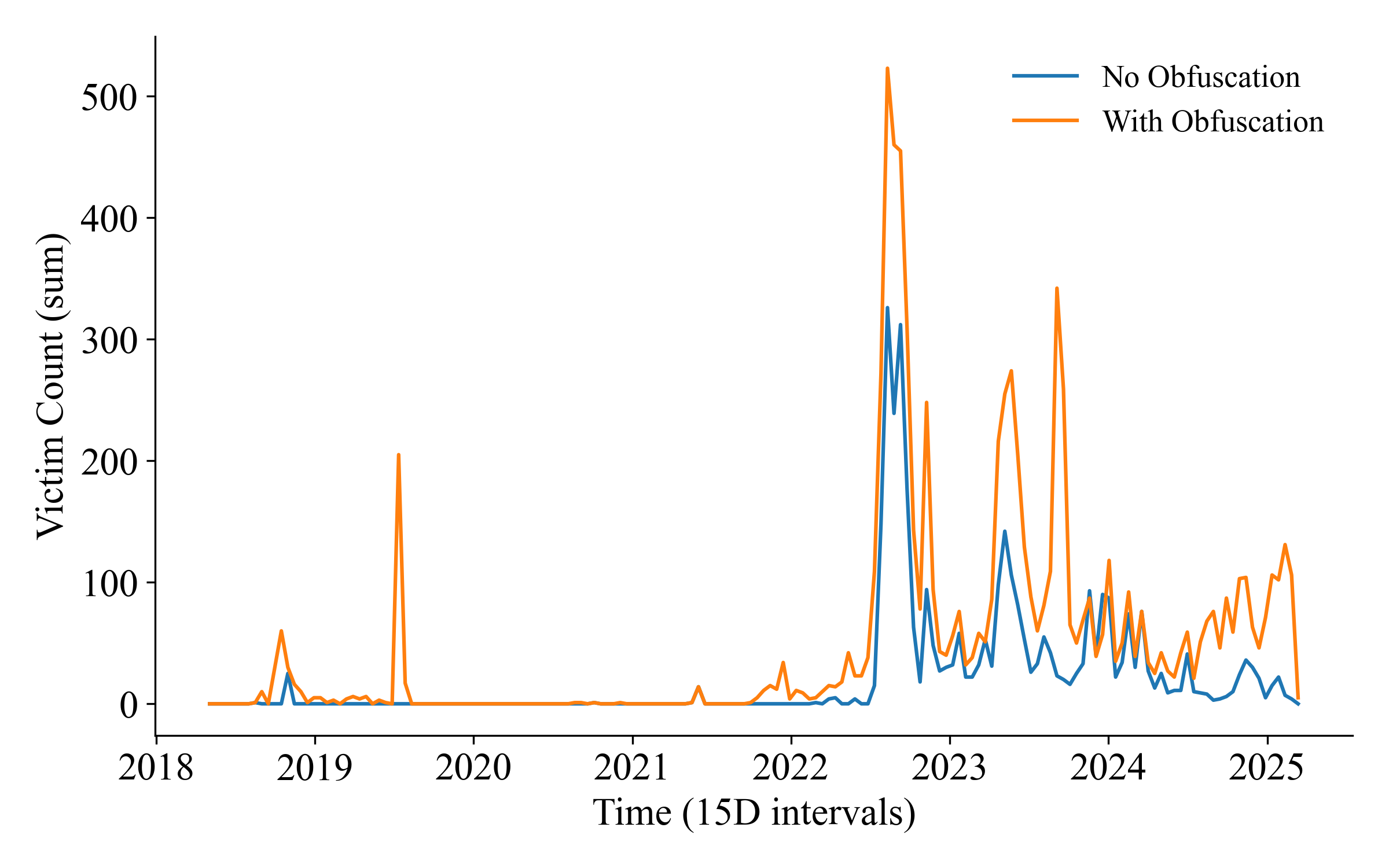}%
        \label{victim_counts}%
    }
    \caption{Time Trend Analysis}
    \vspace{-0.1in}
\end{figure}

\vspace{4pt}\noindent \textbf{Inbound Funds:} Figure~\ref{victim_inbound} shows the inbound funds of each contract group aggregated in the 15-day interval. It can be seen that non-obfuscated contracts (blue line) remain relatively small throughout the period from 2018 to 2025, with occasional spikes. In contrast, the inbound funds of obfuscated contracts (orange line) exhibit multiple peaks, mostly centered between 2022 and 2024. Particularly in June 2019, the \emph{15-day aggregated} inbound funds exceeded 200 ETH. Later in July 2022, the aggregated inbound funds exceeded 250 ETH. Separately, the \emph{single-contract} maximum inbound funds is 201.74 ETH, as reported in Table~\ref{tab:inbound_eth_stats}; this is a different metric from the aggregated curve. Then, in three different months from June 2022 to August 2023, the aggregated inbound funds exceeded 100 ETH each. 

\vspace{4pt}\noindent \textbf{Victim Count.} Figure~\ref{victim_counts} presents the aggregated victim addresses of each contract group in the 15-day interval. It is evident that the number of victims from obfuscated contracts (orange line) consistently exceeds that of non-obfuscated contracts (blue line) throughout the period from 2022 to 2025. Notably, there are sharp increases during specific months (e.g., from the latter half of 2022 to 2023), with the total number of victims within a single 15-day period reaching several hundred. In contrast, the number of victim addresses of non-obfuscated contracts remains much lower during the same period. Such results further highlight that obfuscated contracts can increase the likelihood of deceiving a much larger group of victims.

Scam contracts in the Li2023 dataset often rely on short-lived campaigns and strong incentives to attract participants. When transfer logic is obfuscated, the cost of manual inspection increases, amplifying the impact of these scams.

\rqanswer{3}{Our analysis reveals that obfuscated scam contracts are more active than non-obfuscated ones and have resulted in higher financial losses and a larger number of victims. This finding aligns with our hypothesis: Obfuscation techniques can enable scam contracts to operate more covertly during their initial stages, allowing them to generate greater profits by deceiving more victims.}

\section{Impact on Existing Scam Detection Tools}
\label{tool_detector}
Finally, to answer RQ4, we examine how obfuscation affects the effectiveness of existing malware analysis tools. Note that RQ3 uses the Li2023 scam dataset, while RQ4 evaluates detection on the SourceP Ponzi-labeled set. We run \textsc{SourceP}~\cite{Lu2024}, a representative Ponzi detector, on a \emph{single} dataset: the contract set released with the \textsc{SourceP} paper (summarized in Table~\ref{tab:dataset_classification}). We additionally test MFDPonzi on the same labeled set and observe comparable precision/recall trends, so we report \textsc{SourceP} as the main baseline for clarity. On this dataset, we \textbf{manually inspect and assign} obfuscation labels according to our taxonomy (T1--T7): a contract is labeled \emph{obfuscated} if at least one of T1--T7 is present; otherwise, it is labeled \emph{non-obfuscated}. Ambiguous cases are excluded to avoid label noise. We then compute precision/recall/F1 to quantitatively compare \textsc{SourceP}'s effectiveness on obfuscated vs. non-obfuscated samples.
\begin{table}[t]
\centering
\caption{Effectiveness of SourceP for obfuscated and non-obfuscated Ponzi samples.}
\label{Performance_comparison}
\setlength{\tabcolsep}{6pt}
\renewcommand{\arraystretch}{1.0}
\resizebox{0.95\columnwidth}{!}{%
\begin{tabular}{@{} l r r r r r r @{}}
\toprule
\textbf{Class} & \textbf{Total} & \textbf{TP} & \textbf{FN} & \textbf{Precision} & \textbf{Recall} & \textbf{F1} \\
\midrule
Non-obfuscated & 361 & 287 & 74 & 1.00 & 0.80 & 0.89 \\
Obfuscated     &  92 &  11 & 81 & 1.00 & 0.12 & 0.21 \\
\bottomrule
\end{tabular}
}
\end{table}

\vspace{4pt}\noindent \textbf{Experimental Results.} Table~\ref{Performance_comparison} shows a large gap between non-obfuscated and obfuscated samples. For non-obfuscated contracts, SourceP achieves a precision of 1.00, a recall of 0.80, and an F1 of 0.89, which aligns with its intended capabilities. For obfuscated contracts, recall drops to 0.12 and F1 to 0.21 (precision remains high because the model predicts positives very conservatively), indicating that obfuscation sharply reduces detection effectiveness. High precision coupled with low recall implies conservative behavior: fewer false alarms but far more misses, which is undesirable for early warning.

The pronounced performance disparity between obfuscated and non-obfuscated samples can be attributed to how obfuscation disrupts SourceP's static analysis pipeline. First, externalized transfer paths (T4) relocate the payable sink beyond the intra-contract scope that SourceP primarily analyzes, resulting in under-approximated money-flow graphs and the omission of critical Ponzi indicators. Second, opaque predicates and deep nesting (T2) complicate the control-flow graph, leading to path pruning or premature cut-offs during data and control-flow reconstruction, thereby suppressing essential features on which SourceP relies, such as cyclic payouts and balance-dependent branches. Third, multi-step address synthesis (T1), achieved through hashing, bit-masking, or nonces, disrupts constant propagation and complicates recipient attribution, thereby weakening heuristics that depend on recognizable payout targets. Fourth, dispatcher flattening and function cloning (T5) obscure function boundaries and hinder the reuse of summaries, which degrades feature aggregation at the function level. Fifth, log interference (T6) introduces spurious events and alters the order of informative logs, potentially misleading log-aware heuristics or subsequent sanity checks. Finally, junk code inflation (T7) increases bytecode size and the noise in the intermediate representation, thereby increasing the likelihood of timeouts or the application of conservative defaults. In contrast, non-obfuscated contracts reveal canonical money-flow structures that are consistent with SourceP's feature design and training distribution, resulting in significantly enhanced effectiveness.

We qualitatively inspect representative misclassified obfuscated contracts and find that errors primarily occur in cases involving externalized transfer paths (T4) and path explosion (T2). Additionally, captions and logs often reveal partial analyses, such as truncated call-graph slices. The performance gap between obfuscated and non-obfuscated samples stems from reduced observability of Ponzi-specific signals under obfuscation, rather than from noisy labels or dataset differences. It is essential to note that our evaluation employs an ETH-only victim/flow definition, excluding intermediaries such as routers, centralized exchanges, and maximal extractable value (MEV). This means that the observed gap reflects limitations in our analysis rather than metric contamination.

\rqanswer{4}{Our evaluation results demonstrate that obfuscation has a significant detrimental impact on existing scam detection tools, causing a substantial drop in recall and F1. At a practical level, this observation underscores the importance of developing effective obfuscation analysis techniques to mitigate emerging security threats. }

\section{Discussion}
\label{discussion}
\subsection{Limitations and Threats to Validity}
Our taxonomy (T1--T7) and features (F1--F7) focus on \emph{transfer-path} obfuscation. Within this formal scope, the taxonomy aims for comprehensive coverage. Nonetheless, our work still has the following five limitations: 1) We do not cover ERC-20\slash 721 token flows, bridge\slash mixer interactions, or metamorphic designs via \texttt{CREATE2} or code replacement. These are outside our transfer-path scope and may require additional primitives. 2) The analysis is bytecode\slash IR–only (no dynamic execution or cross-transaction traces). Obfuscation that resolves purely at runtime (e.g., oracle-fed recipients, environment-gated dispatch) can reduce observability and yield false negatives. 3) When the payable path is fully delegated via \texttt{DELEGATECALL}\slash \texttt{CALL} to external contracts, F4 flags externalization but cannot recover the callee’s logic; downstream features (e.g., F1–F7) may thus remain unset. 4) Z-score and absolute guards rely on baseline distributions; different snapshots can shift thresholds slightly (we release seeds to recompute them). 5) Parts of the benchmarks (e.g., the \textsc{SourceP} comparison) use single-annotator manual labels; we exclude ambiguous cases and will release address lists, but residual noise may remain. 6) Our large-scale measurement focuses on the Ethereum mainnet due to its impact and data availability. We leave extending the study to other blockchains (e.g., BSC, Polygon, Arbitrum) for future work. Overall, we regard F1–F7 as a \emph{minimal, extensible core} tailored to the transfer-path scope; extending to non-transfer flows or adding orthogonal runtime evidence is left to future work.

\subsection{Challenges and Future Work}

Obfuscation techniques in smart contracts present significant challenges for auditing and regulatory practices, particularly in scam contracts, MEV bots, and highly centralized systems. These challenges include poor code readability, failure of conventional static detection tools, and delays in regulatory response. Several improvements are necessary to address these issues. Static analysis tools must be enhanced to track deeper control and data flows, while de-obfuscation preprocessing can simplify bytecode for more efficient audits. Dynamic analysis, such as runtime tracing or fuzzing, can bypass superficial obfuscation and validate fund flows during contract execution. Additionally, techniques from traditional software security, such as CFG flattening and semantic normalization, can be applied to EVM bytecode to identify critical logic, such as transfers and permission checks, that obfuscation attempts to conceal. Eventually, establishing collaborative platforms (e.g., a "contract blacklist" or a "high-risk obfuscation" repository) would enable researchers, auditors, and the public to tag suspicious contracts, thereby improving transparency and enhancing collective oversight of the DeFi ecosystem.

\subsection{Obfuscation signals vs. malicious intent}
Our taxonomy (T1--T7) and features (F1--F7) capture \emph{how observable the transfer logic is}, not whether that logic is benign or harmful.
The same primitives often appear in legitimate designs. For instance, T4 can be used in upgradeable proxies and routers, while T5 in flattened dispatchers for gas efficiency. Hence, we treat F1--F7 as \emph{signals} that warrant scrutiny, not verdicts. When we discuss scams, the goal is to \emph{illustrate correlation}, not to claim causation. Disambiguating intent typically requires orthogonal evidence (economic semantics, longitudinal flows, victim-side signals), which is outside the scope of our obfuscation quantification.


\vspace{-0.1in}

\section{Related Work}
\label{related_work}

\vspace{5pt}\noindent\textbf{Classical Obfuscation Techniques.} A large software engineering (SE) literature has studied source/IR-level obfuscation for native/managed code.  Collberg et al.\ offer a widely used taxonomy (layout, control-, data-, and preventive transformations) with potency/resilience/stealth metrics~\cite{Collberg1997Taxonomy}. Representative techniques include opaque predicates and bogus control-flow (control-flow obfuscation), anti-disassembly/anti-decompilation (preventive), instruction substitution and data encoding, and control-flow flattening; industrial-strength implementations such as Obfuscator-LLVM operationalize several of these passes~\cite{LinnDebray2003,LaszloKiss2009CFF,Junod2015SPRO}. 
Surveys and monographs synthesize two decades of progress and limitations~\cite{Schrittwieser2016CSUR,CollbergNagra2009Book}.

\vspace{5pt}\noindent\textbf{Smart Contract Security Analysis.}
Some research employs static analysis to enhance the security and efficiency of smart contracts. USCHUNT \cite{bodell2023proxy} explores the balance between adaptability and security in upgradeable contracts. Madmax~\cite{grech2018madmax} targets vulnerabilities to prevent execution failures, while Slither~\cite{slither:misc} and Smartcheck~\cite{tikhomirov2018smartcheck} automatically detect flaws in Solidity contracts. Symbolic execution is also used to improve security; Mythril~\cite{mythril:misc} analyzes EVM bytecode, EthBMC~\cite{frank2020ethbmc} combines symbolic execution with concrete validation, and Reguard~\cite{liu2018reguard} and Manticore~\cite{manticore:misc} identify reentrancy and other bugs. Smartian~\cite{choi2021smartian} integrates fuzzing with static and dynamic analysis, while Confuzzius~\cite{ferreira2021confuzzius} leverages data dependency insights for fuzzing. ContractFuzzer~\cite{jiang2018contractfuzzer} and Sfuzz~\cite{nguyen2020sfuzz} apply fuzzing to uncover security issues. Research highlights various formal verification methods to enhance the security of smart contracts. Sailfish~\cite{bose2022sailfish} improves state inconsistency detection, while VetSC.\cite{duan2022vetsc} extends DApp verification. Zeus\cite{kalra2018zeus} and Verx~\cite{permenev2020verx} focus on contract safety and condition verification. Smartpulse~\cite{stephens2021smartpulse} analyzes time-based properties, Securify~\cite{tsankov2018securify} identifies security breaches, and Verismart~\cite{so2020verismart} ensures contract safety.

\vspace{5pt}\noindent\textbf{Scam and Ponzi detection.} Prior work has studied Ponzi detection using code-level features and runtime behavior graphs, including SADPonzi~\cite{Chen2021}, SourceP~\cite{Lu2024}, and PonziGuard~\cite{Liang2024}. These approaches are effective on non-obfuscated or lightly obfuscated contracts but are sensitive to transfer-path obfuscation and externalized payout logic. Very recent work such as MFDPonzi~\cite{Cao2025} explores static features from opcode sequences at larger scale; on our labeled set, MFDPonzi shows comparable precision/recall trends to SourceP, so we use SourceP as a representative baseline in the main evaluation.

\vspace{5pt}\noindent\textbf{Advanced Anti-Auditing Techniques.} Recent analyses expose “fake” ownership renunciation: after invoking \texttt{renounceOwnership()}, some contracts zero public fields (e.g., \texttt{owner}, \texttt{getOwner}) yet retain control via concealed state. Shiaeles and Li (2024) document a live case where a benign-looking variable (e.g., \texttt{isTokenReceiver}) stores the deployer’s address, leaving an address-bound backdoor~\cite{Hall2024}. More broadly, innocuous names such as \texttt{isTokenReceiver}, \texttt{failsafe()}, or \texttt{emergency()} can mask administrator-only operations (e.g., privilege restoration or fund extraction).


\vspace{5pt}\noindent\textbf{MEV Bot Obfuscation Techniques.} To protect MEV bots from being front-run by generalized mimicking scripts in the public mempool, practitioners and researchers have developed various obfuscation and privacy-preserving techniques. The most common method is using private relays (e.g., Flashbots) to submit bundles directly to block builders, bypassing the public mempool and preventing adversaries from copying transactions~\cite{yang2024sok,qin2022quantifying}. Intent-based protocols like CoW Swap perform off-chain batch matching of user intents, publishing only the final settlement on-chain to eliminate front-running risks~\cite{yang2024sok}. Gas camouflage techniques, such as locking transactions to specific \texttt{tx.gasprice} values or adding dummy computations, confuse adversarial repricing strategies~\cite{yang2024sok}. Multi-hop contract calls, often paired with flash loans and non-standard swap paths, increase attackers' simulation overhead~\cite{yang2024sok,daian2019flash}. Bytecode-level obfuscations, like inserting \texttt{JUMPI} pseudo-branches or splitting constants via arithmetic, hinder static and dynamic analysis~\cite{yang2024sok}. Recent work has also explored threshold encryption, delayed reveal schemes, and protocol-level MEV “tax” mechanisms to internalize ordering profits~\cite{alnajjar2024mitigating,daian2019flash}. Despite these advancements, systematic research on obfuscation techniques for MEV bots at the smart contract level remains scarce.

\section{Conclusion}
\label{conclusion}
In this paper, we systematically investigate obfuscation techniques in Ethereum smart contracts, providing comprehensive definitions, quantitative methods, and empirical analysis. We introduce seven key quantifiable obfuscation features based on the detailed analysis of transfer instructions. Using a robust Z-score representation model, we analyze the prevalence of obfuscation techniques employed in over 1.04 million Ethereum contracts and conduct an in-depth analysis of 3,000 highly suspicious contracts, revealing four types of malicious contracts. Our further analysis shows that obfuscated scam contracts have a higher financial extraction capability than non-obfuscated scam contracts. We also demonstrate that obfuscation significantly undermines the effectiveness of existing detection tools. At ecosystem scale, obfuscation is routine but risk concentrates in the extreme tail; in practice, auditing should prioritize high-score segments and signals such as F3/F2/F4 for tiered review. Overall, our findings underscore the security risks posed by obfuscation and highlight the urgent need for advanced analytical and detection methodologies to address this evolving threat, enhancing blockchain security and fostering transparency.

\appendices

\newlist{applist}{itemize}{3}
\setlist[applist]{leftmargin=*,labelsep=0.5em,itemsep=2pt,topsep=2pt,parsep=0pt,partopsep=0pt}
\setlist[applist,1]{label=\textbullet}
\setlist[applist,2]{label=\(\circ\)}
\setlist[applist,3]{label=\(\triangleright\)}

\section*{Supplementary Overview}
This supplementary material provides code-level evidence for the three extreme centralization patterns discussed in the main paper (Case IV: Extreme Centralization). We organize the evidence into S1--S3: centralized permission control, arbitrarily adjustable high fee/tax, and lack of genuine fund locking. Each subsection follows a unified structure (summary, code evidence, and tool-visible signals) to improve scanability.

\section{Supplementary Evidence for Extreme Centralization (S1--S3)}
\label{appendix:centralized-cases}
\subsection{S1: Centralized Permission Control}
\label{app:centralized-permission}
\paragraph{Summary.} Such contracts often appear under the guise of "lending protocols", "collateral management", "liquidity safeguarding", etc., mimicking interfaces and function names of well-known protocols (e.g., Compound, Uniswap). However, in reality, they are entirely controlled by a few roles—such as \texttt{admin} and \texttt{liquidateAdmin} (and sometimes more, e.g., \texttt{manager} or \texttt{\_super})—that manage all key operations. 

\begin{applist}
    \item \textbf{Highly Centralized Permissions:} The deployer (Owner) assigns multiple administrative roles to themselves in the constructor, enabling them to modify the oracle, collateral ratios, fee structures, or even forcefully liquidate user assets at any time.
    \item \textbf{Pseudo-"Decentralization":} Although the contract outwardly features multiple roles and safeguard mechanisms, the actual execution authority remains concentrated in a single private key address, leaving users unable to prevent backdoor operations by the owner.
\end{applist}

In practice, these "highly centralized" contracts typically use lengthy, repetitive code and a plethora of events (\texttt{emit}) to create complexity, making it difficult for external auditors to immediately discern their true nature.

\paragraph{Code evidence.} Below, we analyze a real-world case of a contract named \texttt{AegisComptroller.sol} (a pseudonym) to illustrate how such contracts conceal their centralized permission design through role masquerading, numerous redundant functions, and excessive event logging.

\begin{applist}
    \item \textbf{Role Masquerading: Multiple Names, Layered Functions, but Controlled by the Same Address}  
    \textbf{Dual Roles with the Same Private Key:}
    \begin{lstlisting}[language=Solidity]
constructor () public {
  admin = msg.sender;
  liquidateAdmin = msg.sender;
}
\end{lstlisting}
    
    In the constructor, both \texttt{admin} and \texttt{liquidateAdmin} are set to the same address, creating an illusion of multiple roles while, in fact, the same entity controls everything.

    \item \textbf{Redundant Permission Checks and Guard Inflation.} The contract repeats equivalent \texttt{REQUIRE(...)} guards for \texttt{admin} and \texttt{liquidateAdmin}, inflating complexity without adding real protection.

    \item \textbf{Numerous "Administrative" Functions and Spurious Security Checks:}
    \begin{applist}
        \item \textbf{Seemingly Compliant Configuration Functions:}
        \begin{lstlisting}[language=Solidity]
function _setPriceOracle(PriceOracle,
_newOracle) 
public returns (uint) {
    REQUIRE(msg.sender == admin, 
    "SET_PRICE_ORACLE_OWNER_CHECK");
    oracle = _newOracle;
    ...
}
function _setCollateralFactor(
AToken_aToken,
uint _newCollateralFactorMantissa) 
external returns (uint) {
    REQUIRE(msg.sender == admin, 
    "SET_COLLATERAL
    _FACTOR_OWNER_CHECK");
    ...
}\end{lstlisting}
        
        These functions are named very similarly to those in Compound (e.g., "set price oracle" or "set collateral factor"), but they only REQUIRE an admin call and do not incorporate any multisignature or time delay mechanisms.
        \item \textbf{Redundant Role Assignments:}\\
        For example, functions such as \texttt{\seqsplit{\_setMintGuardianPaused()}} , \texttt{\seqsplit{\_setBorrowGuardianPaused()}} , and \texttt{\seqsplit{\_setPauseGuardian()}} ostensibly provide multiple safeguard roles; however, a single admin instruction can control all permissions.
    \end{applist}

    \item \textbf{Direct Backdoor Functions: autoLiquidity / autoClearance}
    \begin{applist}
        \item \textbf{Automated Liquidation Interface:}
        \begin{lstlisting}[language=Solidity]
function autoLiquidity(
address _account, 
uint _liquidityAmount,
uint _liquidateIncome) 
public returns (uint) {
REQUIRE(msg.sender == liquidateAdmin,
"SET_PRICE_ORACLE_OWNER_CHECK");
    ...
// Actually calls 
//autoLiquidityInternal(...)
} \end{lstlisting}
    
    With only the liquidateAdmin (still the deployer's private key), the contract can forcibly seize the collateral of any \_account.
        
        \item \textbf{Internal Forced Transfers:}
        \begin{lstlisting}[language=Solidity]
asset.ownerTransferToken(_owner, 
_account, vars.aTokenBalance);
asset.ownerCompensation(_owner, 
_account, vars.aTokenBorrow);
        \end{lstlisting}
        These functions effectively transfer the user's aToken or lending assets to \texttt{\_owner} (i.e., the administrator).
    \end{applist}

    \item \textbf{Redundant Functions and Events Obscuring the True Process:}
    \begin{applist}
        \item \textbf{Redundant Functions:}  
        Functions such as \texttt{\seqsplit{\_setMintGuardianPaused()}} , \texttt{\seqsplit{\_setBorrowGuardianPaused()}} , \texttt{\seqsplit{\_setTransferPaused()}} , \texttt{\seqsplit{autoLiquidityInternal()}} , and \texttt{\seqsplit{autoClearanceInternal()}} have nearly identical internal logic but are implemented in several different versions.
        \item \textbf{Event Redundancy:}  
        \begin{lstlisting}[language=Solidity]
event AutoLiquidity(address _account, 
uint _actualAmount);
event AutoClearance(address _account, 
uint _liquidateAmount,
uint _actualAmount);
event NewPriceOracle(
PriceOracle _oldPriceOracle, 
PriceOracle _newPriceOracle);
        \end{lstlisting}
        
        The contract defines more than a dozen events, covering actions from market entry and exit to liquidation and oracle switching. The flood of logs during execution makes it difficult for auditors to quickly pinpoint the key backdoor transfers.
    \end{applist}

    \item \textbf{"Guardian" Also Controlled by the Same Admin:}
    \begin{lstlisting}[language=Solidity]
function _setPauseGuardian(
address _newPauseGuardian
) 
public returns (uint) {
    REQUIRE(msg.sender == admin,
    "change not authorized");
    pauseGuardian = _newPauseGuardian;
    ...
}
    \end{lstlisting}
    Although this function appears to assign the pauseGuardian for emergency shutdown of lending/minting, it can still be modified or invoked at any time by the admin (i.e., the same private key), lacking any checks or balances.
\end{applist}

\paragraph{Tool-visible signals.} \begin{applist}
    \item \textbf{Numerous SLOAD/EQ Operations Targeting the Same Owner Storage Slot:}  
    When decompiled or analyzed using SSA, tools will observe that the contract repeatedly reads from the same storage slot (e.g., for \texttt{admin} or \texttt{liquidateAdmin}) and compares it with \texttt{msg.sender}, at a frequency far exceeding that of typical contracts.
    
    \item \textbf{Backdoor Functions Dependent on External Calls:}  
    CALL instructions such as ownerTransferToken(...) and ownerCompensation(...) may appear in multiple locations and are controlled by a single address, indicating that the fund flow ultimately converges to the same external address.
    
    \item \textbf{High Function Redundancy and Excessive \texttt{emit} Usage:}  
    Analysis of the control flow graph (CFG) or branch structure reveals multiple function blocks with extremely high similarity, and multiple \texttt{emit} events appear before and after the \texttt{Transfer}. This results in an unusually high proportion of redundant instructions.
\end{applist}

In summary, contracts employing "highly centralized permission control" create audit noise through techniques such as role name masquerading, dispersed configuration functions, and excessive event logging. Yet, all critical operations remain controlled by a single address, clearly posing a Rug Pull risk.

\subsection{S2: Unreasonable and Arbitrarily Adjustable High Fee / Tax Contracts}
\label{app:high-fee-tax}

\paragraph{Summary.} Such contracts typically adopt a "token issuance + Automated Market Maker (AMM)" model. They claim to offer various functions such as liquidity management, charity funds, and marketing pools, but their true purpose is to harvest ordinary users by imposing exorbitant and arbitrarily adjustable "fees" or "taxes." Their main characteristics include:

\begin{applist}
    \item \textbf{Exorbitant Fee Rates:}  
    The fee rates for buying, selling, or withdrawing can often range from 10\% to 50\%, and may even be instantly adjusted up to 99\%, far exceeding normal transaction fees.
    
    \item \textbf{Multiple Nominal Tax Categories:}  
    Contracts often declare several tax types (e.g., "Marketing Tax", "Liquidity Tax", "Development Tax"), yet the funds ultimately flow to a single EOA (the project’s address).
    
    \item \textbf{Arbitrarily Adjustable:}  
    Through functions like setTaxes() or similar, the contract administrator (Owner) can increase the fee rate from as low as 3\% to as high as 99\% at any time, without requiring any voting, multisignature, or delay. Consequently, users may unexpectedly face exorbitant fees, and a substantial amount of funds flows directly into the project’s wallet.
    
    \item \textbf{Redundant Event Obfuscation:}  
    A large number of events (e.g., \texttt{FeeEvent}, \texttt{logTax}, or other unrelated logs) are inserted before and after critical transfers or transactions, masquerading as "transparent operations." In reality, these merely serve to conceal the true harvesting logic, making it difficult for auditors or users to quickly discern the actual fund flow.
\end{applist}

In summary, such contracts leverage a "high liquidity + high tax" structure to attract initial funds, and once the token gains popularity, they can instantly raise the fee rate or even lock transactions, resulting in heavy losses for users while the project continuously profits.

\paragraph{Code evidence.} Below, we use the "GATSOKU" contract as an example to illustrate the typical implementations in this type of scam contract with respect to high fee rates, on-demand adjustability, and multiple event obfuscations.

\begin{applist}
    \item \textbf{Exaggerated Tax Rate Settings and On-Demand Adjustments:}
    \begin{applist}
        \item \textbf{Initial High Tax:}
        \begin{lstlisting}[language=Solidity]
uint256 public taxForLiquidity = 47;
uint256 public 
taxForMarketingHostingDevelopment 
= 47;
        \end{lstlisting}
        At deployment, the contract sets a transaction tax rate of 47\% + 47\% = 94\%, which can easily be raised to 99\%.
        
        \item \textbf{Temporary Adjustments:}
        \begin{lstlisting}[language=Solidity]
function postLaunch() 
external onlyOwner {
    taxForLiquidity = 0;
    taxForMarketingHostingDevelopment 
    = 3;
    ...
}
        \end{lstlisting}
        With the \texttt{onlyOwner} modifier, the administrator can instantly adjust the tax rates without any multisignature or delay.
    \end{applist}
    
    \item \textbf{All Taxes Consolidated to a Single Address, with No Lockup or Custody:}
    \begin{lstlisting}[language=Solidity]
address public marketingWallet
= 0x02796bAeb663......;
bool sent = 
payable(marketingWallet).send(
    address(this).balance
);
REQUIRE(sent, "Failed to send ETH");
    \end{lstlisting}
    After taxation, all funds are transferred to marketingWallet, which the administrator can change at any time. There is no external custody or lockup, nor any community oversight mechanism.
    
    \item \textbf{Complex Fee Calculations and Numerous Auxiliary Functions During Transactions:}
    \begin{applist}
        \item \textbf{Core \texttt{\_transfer()} Function:}
        \begin{lstlisting}[language=Solidity]
function _transfer(address from, 
address to,
uint256 amount) 
internal override 
{
...
if ((from == uniswapV2Pair 
|| to == uniswapV2Pair)
&& 
!inSwapAndLiquify) {
   if (!_isExcludedFromFee[from] 
   && !_isExcludedFromFee[to]) {
     uint256 marketingShare = 
     (amount * taxForMarketingHosting
     Development) 
     / 100;
     uint256 liquidityShare = 
     (amount * taxForLiquidity) / 100;
     // Transfer the tax portion to 
     //this contract, 
     //then later to marketingWallet
     super._transfer(from, address(this),
     (marketingShare + liquidityShare));
      _marketingReserves += marketingShare;
   }
 }
  super._transfer(from, to, 
  transferAmount);
}
        \end{lstlisting}
        The tax portion is continuously retained within the contract and eventually transferred to marketingWallet.
        
        \item \textbf{Complex Swap/Liquify Functions:}
        \begin{lstlisting}[language=Solidity]
function _swapTokensForEth(
uint256 tokenAmount
) 
private lockTheSwap 
{
    ...
    uniswapV2Router.
    swapExactTokensForETHSupporting
    FeeOnTransferTokens(
        tokenAmount,
        0, 
        path,
        address(this),
        block.timestamp
    );
}
function _addLiquidity(
uint256 tokenAmount,
uint256 ethAmount) 
private lockTheSwap {
    uniswapV2Router.addLiquidityETH{
    value: ethAmount
    }(
        address(this),
        tokenAmount, 
        0,
        0,
        marketingWallet,
        block.timestamp
    );
}
        \end{lstlisting}
        These functions increase the complexity of the audit, giving the impression of professional automated market-making logic, though ultimately a large amount of funds still flows to a single address.
    \end{applist}
    
    \item \textbf{Redundant Event Insertion and "Unlock Function" Disguising:}  
    The code also defines events and structures that are completely unrelated to taxation, such as \texttt{UserUnlocked} and \texttt{ChannelUnlocked}:
    \begin{lstlisting}[language=Solidity]
struct userUnlock {
string tgUserName;
bool unlocked;
... }
event UserUnlocked(
string tg_username,
uint256 unlockTime
);
    \end{lstlisting}
    Such unrelated logic is dispersed throughout the code, increasing the difficulty of reading and auditing, and thereby obscuring the core tax-harvesting operations.
\end{applist}

\paragraph{Tool-visible signals.} \begin{applist}
    \item \textbf{High Complexity in Transfer Logic.} Within the \texttt{\seqsplit{\_transfer()}} function, frequent string operations and branch conditions elevate the branch-depth and log-density signals.
    
    \item \textbf{External CALL Tracing:}  
    After taxation, external contracts (e.g., \texttt{uniswapV2Router}) are often called to perform token swaps, and the resulting ETH is sent to the project's address. Tools can detect this via backward slicing—when the owner arbitrarily changes variables, the tax rate takes effect immediately, marking it as a high-risk feature.
    
    \item \textbf{Abundant Irrelevant Events or States:}  
    Irrelevant events (such as \texttt{UserUnlocked} or \texttt{CostUpdated}) frequently occur before and after the Transfer, which tools can flag as "log noise" or "potential obfuscation techniques."
\end{applist}

Overall, while these contracts superficially implement "automated liquidity management" and insert functions and events unrelated to taxation, their core logic remains that the administrator can instantaneously raise the fee rate and harvest funds from retail investors. Once the tax rate increases to 90\%–99\%, ordinary users can hardly liquidate their assets, and their funds are continuously funneled into the project’s private pocket.

\subsection{S3: Contracts Without Genuine Fund Locking}
\label{app:no-locking}
\paragraph{Summary.} Contracts of this type typically attract users by advertising themselves as “DeFi Farms / Staking / NFT Pools / Lending” platforms, promising high yields or robust security measures. However, their fundamental characteristics are as follows:
\begin{applist}
    \item The contract does not actually lock user funds in a decentralized manner.
    \item The Owner possesses a backdoor that allows funds to be transferred or drained at any time.
    \item Functions such as \texttt{\seqsplit{emergencyWithdraw()}} , \texttt{\seqsplit{emergencyEnd()}} , or \texttt{\seqsplit{emergencyRescue()}} are exclusively available to the project team, leaving ordinary users defenseless.
\end{applist}

Once users deposit funds into the contract, their money appears to be “staked” or “custodied” in a “Bank” or “Strategy.” In reality, a single Owner key is sufficient to withdraw the funds instantly. The long functions and complex data structures (e.g., multiple layers of strategy, Bank, Deposit) significantly increase the difficulty of auditing, thereby concealing the true centralized backdoor logic.

\paragraph{Code evidence.} Taking \texttt{Staking.sol} as an example, we illustrate how these contracts mislead outsiders with complex “strategy management,” “emergency withdrawals,” and “cross-contract calls,” while in reality allowing the Owner to control all assets.

\textbf{Lack of Genuine "Locking" of Liquidity and Strategies:}

\textbf{Bank \& Strategy:}  
The contract defines data structures such as \texttt{Bank}, \texttt{StrategyParameters}, and \texttt{Deposit} to record strategy names, staked amounts, safety flags, etc. At first glance, user funds appear to be systematically custodied and yield calculated:
\begin{lstlisting}[language=Solidity]
struct StrategyParameters {
    string name;
    bool isSafe;
    uint256 rateX1000;
    bool isPaused;
    uint256 withdrawId;
}
function purchaseStableTokens(
string memory strategyName,
uint256 amount)
    external
    onlyOwner
{
    REQUIRE(amount > 0, 'amount = 0');
    REQUIRE(strategiesParameters[strategyName]
    .rateX1000 != 0, 
    'Strategy is not exist');
    _stableToken.safeTransferFrom(
    msg.sender, 
    address(this), amount);
    stableTokenBank[strategyName]
    += amount;
    ...
    emit AddBank(
    block.timestamp,
    strategyName,
    amount);
}
\end{lstlisting}
However, the funds ultimately remain under the contract’s control, and are freely managed by functions guarded by \texttt{onlyOwner}, without any multisignature or time delay.

\textbf{Fake Process:}  
Some functions (e.g., \texttt{\seqsplit{requestWithdraw(...)}} and others) appear to REQUIRE user initiation, but the key steps or conditions can be forcefully modified by the Owner. For example:
\begin{lstlisting}[language=Solidity]
function requestWithdraw(
uint256 depositId)
external 
...
{
    ...
    if (_withdrawFromBank(depositId)) {
        return;
    }
    ...
}
\end{lstlisting}
If the project inserts additional conditions or backdoor calls in \texttt{\seqsplit{\_withdrawFromBank(...)}} then any “locking” restrictions can be bypassed.

\textbf{"Emergency/Backend" Functions for On-Demand Withdrawals:}

\textbf{Claiming to "Protect Users":}  
Contracts often claim in their documentation that in the event of a security incident, functions such as \texttt{\seqsplit{emergencyWithdraw()}} or \texttt{\seqsplit{fulfillDeposited(...)}} can be activated to protect users. In the code, however, these functions are mostly restricted to onlyOwner, with no multisignature or community approval:
\begin{lstlisting}[language=Solidity]
function claimTokens(
uint256 maxStableAmount
) 
external onlyOwner 
{
    // Convert user deposits to 
    //stableToken and then transfer
    //to msg.sender (Owner)
    _stableToken.safeTransfer(
    msg.sender, stableAmount);
    ...
}
\end{lstlisting}
Although users might still see “balance = 100” in the internal ledger, the actual funds have long been withdrawn.

\textbf{Multiple Fulfill Interfaces:}
\begin{lstlisting}[language=Solidity]
function fulfillDeposited(
string memory strategyName,
uint256 amountMaxInStable
) 
external onlyOwner {
... 
}
function fulfillRewards(
string memory strategyName,
uint256 amountMaxInStable
) 
external onlyOwner 
{
... 
}
\end{lstlisting}
Under the guise of “liquidation” or “reward,” these functions actually serve as backdoor withdrawal mechanisms. When combined with delegatecall to an external contract (e.g., \texttt{StakingShadow}), the obfuscation is further deepened.

\textbf{Redundant/Highly Similar Functions:}

Multiple withdrawal/transfer functions such as \texttt{\seqsplit{\_withdrawFromBank(...)}} , \texttt{\seqsplit{withdraw(...)}} , \texttt{\seqsplit{fulfillDeposited(...)}} , and \texttt{\seqsplit{claimTokens(...)}} —despite having different names, share similar logic and can all be used to extract or transfer assets.
\begin{lstlisting}[language=Solidity]
function _withdrawFromBank(
uint256 depositId
) 
internal ... 
{
    ...
    _claim(depositId);
    ...
    emit Withdrawed(
    block.timestamp,
    depositId);
}
\end{lstlisting}
\textbf{Splitting into External Contracts:}  
Subcontracts like StakingShadow are employed to offload part of the logic via delegatecall. Although they appear to separate some functionality from the main contract, they ultimately merge at runtime to form a unified permission chain.

\paragraph{Tool-visible signals.} \begin{applist}
    \item \textbf{Function Similarity Analysis:}  
    Automatic detection of functions such as \texttt{\seqsplit{\_withdrawFromBank}} , \texttt{\seqsplit{withdraw}} , \texttt{\seqsplit{claimTokens}} , etc., often reveals highly similar instruction or control flow patterns, indicating redundant withdrawal logic.
    
    \item \textbf{Abundant External CALLs and Owner Dependency:}  
    External calls such as \texttt{\seqsplit{functionDelegateCall(...)}} , \texttt{\seqsplit{\_stableToken.safeTransfer(...)}} , and \texttt{\seqsplit{\_router.swapExactTokensForTokens(...)}} are all subject to onlyOwner control, showing that ultimate control over funds is extremely centralized.
    
    \item \textbf{Lack of Locking/Multisignature:}  
    Tools can observe that there are no multisignature or delayed execution functions, implying that the so-called “Staking” or “Liquidity Pool” does not actually prevent the Owner from transferring funds at any time.
\end{applist}

In summary, these contracts, through carefully designed multi-layer data structures and extensive function wrappers, disguise seemingly complex “staking/mining/yield management” as a closed backdoor. While users only see attractive yield figures on the front end, they cannot prevent the Owner from withdrawing funds at will, potentially resulting in a rug pull or a situation where funds become unrecoverable.

\bibliographystyle{IEEEtran}
\bibliography{refs}

\end{document}